\documentclass[12pt]{article}
\usepackage{epsf}
\usepackage[dvips]{graphicx,psfrag}

\renewcommand{\thefootnote}{\#\arabic{footnote}}
\renewcommand{\theequation}{\thesection.\arabic{equation}}

\begin{document}

\newcommand{\gtrsim}{ \mathop{}_{\textstyle \sim}^{\textstyle >} }
\newcommand{\lesssim}{ \mathop{}_{\textstyle \sim}^{\textstyle <} }

\renewcommand{\thefootnote}{\fnsymbol{footnote}}
\setcounter{footnote}{0}
\begin{titlepage}

\def\thefootnote{\fnsymbol{footnote}}

\begin{center}

\hfill TU-654\\
\hfill hep-ph/0206026\\
\hfill June, 2002\\

\vskip .5in

{\Large \bf

Cosmic Density Perturbations \\ from Late-Decaying
Scalar Condensations

}

\vskip .45in

{\large
Takeo Moroi and Tomo Takahashi
}

\vskip .45in

{\em
Department of Physics, Tohoku University, Sendai 980-8578, Japan
}

\end{center}

\vskip .4in

\begin{abstract}

    We study the cosmic density perturbations induced from fluctuation
    of the amplitude of late-decaying scalar condensations (called
    $\phi$) in the scenario where the scalar field $\phi$ once
    dominates the universe.  In such a scenario, the cosmic microwave
    background (CMB) radiation originates to decay products of the
    scalar condensation and hence its anisotropy is affected by the
    fluctuation of $\phi$.  It is shown that the present cosmic
    density perturbations can be dominantly induced from the
    primordial fluctuation of $\phi$, not from the fluctuation of the
    inflaton field.  This scenario may change constraints on the
    source of the density perturbations, like inflation.  In addition,
    a correlated mixture of adiabatic and isocurvature perturbations
    may arise in such a scenario; possible signals in the CMB power
    spectrum are discussed.  We also show that the simplest scenario
    of generating the cosmic density perturbations only from the
    primordial fluctuation of $\phi$ (i.e., so-called ``curvaton''
    scenario) is severely constrained by the current measurements of
    the CMB angular power spectrum if correlated mixture of the
    adiabatic and isocurvature perturbations are generated.

\end{abstract}

\end{titlepage}

\renewcommand{\thepage}{\arabic{page}}
\setcounter{page}{1}
\renewcommand{\thefootnote}{\#\arabic{footnote}}
\setcounter{footnote}{0}

\renewcommand{\theequation}{\thesection.\arabic{equation}}

\section{Introduction}
\label{sec:introduction}
\setcounter{equation}{0}

In the recent years, observation of the cosmic microwave background
(CMB) anisotropy has been greatly improved.  After the discovery of
the CMB anisotropy by COBE \cite{APJ464L1} at the angular scale
$\theta\gtrsim 7^\circ$, there have been many efforts to improve the
measurements.  In particular, recent balloon and ground-based
experiments observed the CMB anisotropy at smaller angular scale of
$\theta\sim O(0.1^\circ)$ \cite{aph0104460,aph0104459,aph0104489}.
Then, the observation of the CMB anisotropy is expected to be greatly
improved by the on-going and future satellite experiments, MAP
\cite{MAP} and PLANCK \cite{PLANCK}; after these experiments, the CMB
angular power spectrum $C_l$ will be determined at $O(1\ \%)$ level up
to the multipole $l\lesssim 1000-2000$.

With these measurements, our understanding of the evolution of the
universe is also being improved.  Importantly, since the CMB power
spectrum is sensitive to origins and evolutions of the cosmic density
perturbations (as well as to the cosmological parameters), we are now
able to constrain scenarios of generating the cosmic density
perturbations.

Among various scenarios, inflation \cite{PRD23-347} is probably the
most popular and well-motivated one to provide the source of the
cosmic density perturbations.  In the simplest case, all the
components in the universe (other than the cosmological constant),
like the photon, baryon, cold dark matter (CDM), and so on, originate
to decay products of the inflaton field $\chi$ which is the scalar
field responsible for the inflation.  The inflaton field fluctuates
during the inflation and it becomes the source of the cosmic density
perturbations.  One of the most important consequences of such a
scenario is that, since all the components in the universe are
produced from the inflaton field, there is no entropy perturbation
between any of two components and hence the density fluctuations
become adiabatic.  Importantly, assuming the standard evolution of the
universe after the inflation and using a reasonable set of the
cosmological parameters, it is now widely believed that the observed
CMB power spectrum is consistent with the one predicted from the
scale-invariant spectrum which can arise from some class of inflation
models \cite{hep-ph_0201264}.

From the particle-physics point of view, however, the simplest
scenario may not be the case and there exist possible sources of the
cosmic density perturbations other than the inflaton field.  In
particular, in various scenarios, light scalar fields are introduced
which dominate the universe at early epochs.  (We denote such a scalar
field as $\phi$.) For example, in the scenario of Affleck-Dine
baryogenesis \cite{NPB249-361}, condensation of the squark and slepton
fields is converted to the baryon-number asymmetry of the universe.
If this is the case, there may exist an epoch when the universe is
dominated by the Affleck-Dine field.  Then, at the time of the decay
of the Affleck-Dine field, the late-time entropy production occurs.
Other candidate is flat directions in the theory space.  In
particular, in the superstring theory \cite{Polchinski}, there are
various flat directions parameterized by scalar fields called moduli
fields.  Such moduli fields are expected to acquire masses from the
effect of the supersymmetry breaking and hence their masses can be
much lighter than the Planck scale.  If the initial amplitudes of the
moduli fields are large, the universe is once dominated by the moduli
field and is reheated at the time of the decay of the moduli fields.
(See Refs.\ \cite{PRL131-59,heavy-moduli} for the scenario with
cosmological moduli fields.)  In addition, in Refs.\ 
\cite{snu-leptogen} it is pointed out that the baryon-number asymmetry
of the universe can be explained if the universe was once dominated by
the condensation of the right-handed scalar neutrino.  In this case,
the universe also experiences the late-time entropy production.  In
addition, axion-like scalar field is proposed as a seed of the cosmic
density perturbations \cite{axion=seeds} in the pre-big-bang
\cite{PreBigBang} and the ekpyrotic \cite{ekpyrotic} scenarios.

Such scalar fields may acquire fluctuations of the amplitude in the
early universe (in particular, during the inflation).  Then, the
scalar field eventually decays and reheats the universe.  Importantly,
in this class of scenario, the CMB radiation we observe today
originates to $\phi$ rather than to the inflaton.  As a result, one
can expect that the fluctuation of the scalar-field amplitude affects
the cosmic density perturbations
\cite{PRD42-313,NPB626-395,PLB524-5,PLB522-215}.  In particular, in
such a scenario, adiabatic and isocurvature perturbations may be
generated with cross-correlation.\footnote
{For other mechanisms of generating correlated mixture of the
adiabatic and isocurvature perturbations, see \cite{adi-iso}.}
Indeed, in Ref.\ \cite{PLB522-215}, the CMB angular power spectrum is
calculated in such a framework, and it was shown that effects of the
correlated isocurvature perturbation may be large enough to be seen in
the on-going and future experiments.  In addition, it is possible that
the dominant part of the cosmic density perturbations observed today
may originate to the primordial perturbation in the amplitude of
$\phi$ rather than to the fluctuation of the inflaton amplitude.
(This kind of mechanism is sometimes called ``curvaton mechanism.'')

In this paper, we consider effects of such scalar-field condensations
to the cosmic density perturbations without relying on any particular
scenarios.  Assuming that the universe was once dominated by the
scalar-field condensation and that the reheating occurred at a later
stage with generating large amount of entropy, we study the evolutions
of the cosmic density perturbations and consider the CMB angular power
spectrum.  As we will see, the CMB power spectrum can be affected in
various ways depending on scenarios.  Interestingly, if the decay
product of the $\phi$ field does not generate the baryon asymmetry or
the CDM component, then the adiabatic and isocurvature perturbations
can be both generated with cross-correlation.  In this case, the CMB
power spectrum may significantly change its behavior from the
adiabatic result.  In particular, the simple ``curvaton'' scenario, in
which the cosmic density perturbations are generated only from the
primordial fluctuation of the amplitude of $\phi$, is severely
constrained by the current observations of the CMB angular power
spectrum {\sl if the adiabatic and isocurvature perturbations are
generated with cross-correlation}.

The organization of the rest of this paper is as follows.  In Section
\ref{sec:scenario}, we introduce the scenario we consider and follow
the thermal history of the universe.  Evolutions of the cosmic density
perturbations in such a scenario is discussed in Section
\ref{sec:evolution}.  In particular, it is discussed how the density
perturbation in the radiation is affected by the primordial
fluctuation of the amplitude of $\phi$.  Then, in Section
\ref{sec:cl}, we overview the behavior of the CMB anisotropy in our
scenario.  Detailed discussions on the CMB power spectrum for the
cases with and without the isocurvature perturbations are given in
Sections \ref{sec:noentropy} and \ref{sec:entropy}, respectively.
Section \ref{sec:conclusion} is devoted for the conclusion.

\section{Scenario}
\label{sec:scenario}
\setcounter{equation}{0}

Let us first introduce the scenario we have in mind.  In this paper,
we consider the thermal history with late-time entropy production
which is due to the decay of the scalar field $\phi$.

For our study, it is important to understand the behavior of the
scalar field in the expanding universe.  For this purpose, we start
with the Lagrangian of the form
\begin{eqnarray}
    {\cal L} = \frac{1}{2} g^{\mu\nu} \partial_\mu\phi\partial_\nu\phi
    - V(\phi),
\end{eqnarray}
where $V(\phi)$ is the potential of the scalar field.  In our
discussion, we adopt the parabolic potential for the scalar field;
\begin{eqnarray}
V(\phi)=\frac{1}{2} m_\phi^2 \phi^2.
\end{eqnarray}
From this Lagrangian, we can derive the energy density $\rho_\phi$ and
the pressure $p_\phi$ of the scalar field condensation; for the
zero-mode, we obtain
\begin{eqnarray}
    \rho_\phi = 
    \frac{1}{2} \dot{\phi}^2 + \frac{1}{2} m_\phi^2 \phi^2,~~~
    p_\phi = \frac{1}{2} \dot{\phi}^2 - \frac{1}{2} m_\phi^2 \phi^2,
\end{eqnarray}
where the ``dot'' denotes the derivative with respect to time $t$.

In addition, in the expanding universe, equation of motion of the
scalar field is given by
\begin{eqnarray}
    \ddot{\phi} + 3 H \dot{\phi} + m_\phi^2 \phi = 0,
\end{eqnarray}
where $H\equiv\dot{a}/a$ is the expansion rate with $a$ being the
scale factor.  The behavior of $\phi$ depends on the relative size of
the expansion rate $H$ and the scalar mass $m_\phi$.  When the
expansion rate is much larger than $m_\phi$, the second term, which
plays the role of the friction term, is significant and $\dot{\phi}$
(almost) vanishes.  In this case, change of the scalar amplitude can
be neglected and $\phi$ takes a constant value.  We call this epoch as
``slow-roll'' epoch since $\dot{\phi}$ is negligibly small.  On the
contrary, when $H\lesssim m_\phi$, the scalar field oscillates around
the minimum of the potential.  In particular, when $H\ll m_\phi$, the
oscillation of the scalar field becomes fast enough so that the
relation $\frac{1}{2}\langle\dot{\phi}^2\rangle_{\rm osc}
=\frac{1}{2}m_\phi^2\langle\phi^2\rangle_{\rm osc}$ holds, where
$\langle\cdots\rangle_{\rm osc}$ denotes the average over the time
scale much longer than $m_\phi^{-1}$ (but shorter than $H^{-1}$).
With this relation, the energy density and the pressure of the scalar
field are given as
\begin{eqnarray}
    \langle \rho_\phi \rangle_{\rm osc} = 
    \left\langle 
        \frac{1}{2}\dot{\phi}^2 + \frac{1}{2} m_\phi^2 \phi^2 
    \right\rangle_{\rm osc} = 
    m_\phi^2 \langle \phi^2 \rangle_{\rm osc},~~~
    \langle p_\phi \rangle_{\rm osc} = 
    \left\langle 
        \frac{1}{2}\dot{\phi}^2 - \frac{1}{2} m_\phi^2 \phi^2 
    \right\rangle_{\rm osc} = 0.
\end{eqnarray}
As we can see, the pressure of the scalar field effectively vanishes
at this epoch and the equation-of-state parameter for the scalar field
becomes $\omega_\phi=0$.  In this case, the energy density of the
scalar field behaves as that of the non-relativistic component.
Consequently, when $\Gamma_\phi\ll H\ll m_\phi$ (with $\Gamma_\phi$
being the decay rate of $\phi$), $\rho_\phi$ decreases as $a^{-3}$.

Now we discuss the thermal history of the universe with the scalar
field $\phi$.  In particular, we consider the case where the initial
amplitude of the scalar field $\phi_{\rm init}$ is non-vanishing so
that $\phi$ may become cosmologically significant.  With such an
initial condition, let us follow the evolution of the universe
starting with the inflationary period.

As in the conventional cases, we adopt the epoch of (slow-roll)
inflation in the early stage.  As well as solving the horizon and
flatness problems, inflation provides the source of the cosmic density
perturbations.  We assume that the mass of the scalar field $\phi$ is
much smaller than the expansion rate of the universe during the
inflation $H_{\rm inf}$.  If this is the case, the energy density of
the scalar field is much smaller than the total energy density of the
universe as far as the amplitude of the scalar field is smaller than
the Planck scale.  Then, the potential of the scalar field $\phi$ is
effectively flat during inflation and there is no strong reason to
have vanishing initial amplitude of $\phi$.\footnote
{If the minimum of the potential of $\phi$ is a symmetry-enhanced
point, however, initial value of $\phi$ may vanish \cite{DinRanTho}.}
In addition, if $m_\phi\ll H_{\rm inf}$, $\phi$ is slow-rolling during
inflation and it keeps its initial amplitude.  Thus, we assume a
non-vanishing value of the initial amplitude $\phi_{\rm init}$.

In the slow-roll inflation models, the inflaton field $\chi$
oscillates after the inflation and decays.  Then, the universe is
reheated.  (We call this epoch as the ``first reheating'' since, in
the scenarios we consider, the universe is reheated again due to the
decay of $\phi$.)  The energy density of the inflaton field also
decreases as $a^{-3}$ in this oscillating period, and hence the
expansion rate of the universe is being decreased.  Then, at the time
when the expansion rate of the universe becomes comparable to the
decay rate of the inflaton $\Gamma_\chi$, the inflaton decays and
reheats the universe.  Equating the expansion rate and the decay rate
of the inflaton field, the reheating temperature is estimated to be
\begin{eqnarray}
    T_{\rm R1} \sim g_{*}^{-1/4} \sqrt{\Gamma_\chi M_*},
\end{eqnarray}
where $g_*\sim O(100)$ is the effective number of the massless degrees
of freedom,\footnote
{In this section, we neglect the temperature dependence of $g_*$ since
it does not significantly change the following discussion.}
$M_*\simeq 2.4\times 10^{18}\ {\rm GeV}$ is the reduced Planck scale,
and we adopt the approximation of the instantaneous reheating at the
time of $H=\Gamma_\chi$.  In addition, here (and in the following
order-of-estimations in this section), we neglect numerical constants
of $O(1)$.

Evolution of the scalar field $\phi$ depends on the relative size of
$m_\phi$ and $\Gamma_\chi$.  If the mass $m_\phi$ is smaller than
$\Gamma_\chi$, $\phi$ starts its oscillation after the decay of the
inflaton; otherwise, $\phi$ starts to move while the inflaton field is
still oscillating.  In studying the thermal history with the scalar
field $\phi$, it is also important to specify the initial value of the
scalar field $\phi_{\rm init}$.  When $\phi$ starts to move, $H\sim
O(m_\phi)$ and hence the total energy density of the universe is
$O(m_\phi^2M_*^2)$.  Thus, if $\phi_{\rm init}$ is smaller than $M_*$,
the scalar field is a sub-dominant component when it starts to move,
and hence the evolution of the universe at this epoch is controlled by
other components, oscillating inflaton field or radiation.  If
$\phi_{\rm init}\sim M_*$, on the contrary, energy density of $\phi$
becomes comparable to the total energy density.  In this case, the
scalar field $\phi$ never becomes minor component until it decays.

Let us first consider the case where $T_{\rm R1}$ is high enough so
that $\phi$ starts to oscillate after the decay of the inflaton.  This
happens when $m_\phi\lesssim\Gamma_\chi$.  In this case, the scalar
field starts to oscillate at the temperature
\begin{eqnarray}
    T_{\rm osc} \sim g_*^{-1/4} \sqrt{m_\phi M_*}.
\end{eqnarray}
After this epoch, the energy density of the scalar field $\rho_\phi$
decreases as $a^{-3}$ while that of radiation is proportional to
$a^{-4}$, and hence the energy density of the scalar field decreases
slower than that of the radiation.  Thus, when the amplitude of the
scalar field is as large as $\phi_{\rm eq}\sim\phi_{\rm
init}^4/M_*^3$, the energy density of the scalar field becomes
comparable to that of radiation if the scalar field does not decay
before this epoch.  Expansion rate at this epoch is $H_{\rm eq}\sim
(\phi_{\rm init}^4/M_*^4)m_\phi$, and $H_{\rm eq}$ becomes larger than
the decay rate of $\phi$ if
\begin{eqnarray}
    \phi_{\rm init} \gtrsim  (\Gamma_\phi / m_\phi)^{1/4} M_*
    \sim g_*^{1/8} (T_{\rm R2}^2/m_\phi M_*)^{1/4} M_*
    ~~~:~~~ m_\phi <\Gamma_\chi,
    \label{phimin}
\end{eqnarray}
where $T_{\rm R2}$ is the reheating temperature after the decay of
$\phi$ and is estimated as
\begin{eqnarray}
    T_{\rm R2} \sim g_*^{-1/4} \sqrt{\Gamma_\phi M_*}.
\end{eqnarray}
If the condition (\ref{phimin}) is satisfied, the scalar field
dominates the universe before it decays.

If the reheating temperature after the inflation $T_{\rm R1}$ becomes
low, the scalar field $\phi$ may start to oscillate before the
inflaton field decays; this is the case if $\Gamma_\chi\lesssim
m_\phi$.  In this case, when the inflaton field decays, the amplitude
of the scalar field is $(\Gamma_\chi/m_\phi)\phi_{\rm init}$.  Then,
the energy densities of $\phi$ and radiation become comparable when
$\phi\sim (\Gamma_\chi/m_\phi)\phi_{\rm init}^4/M_*^3$, if $\phi$ does
not decay before its amplitude becomes smaller than this value.  Then,
the condition to have a $\phi$-dominated epoch is
\begin{eqnarray}
    \phi_{\rm init} \gtrsim (\Gamma_\phi /\Gamma_\chi)^{1/4} M_*
    \sim ( T_{\rm R2}/T_{\rm R1} )^{1/2} M_*
    ~~~:~~~ m_\phi > \Gamma_\chi.
\end{eqnarray}

If the initial amplitude of $\phi$ becomes larger than $\sim M_*$, on
the other hand, the $\phi$ field slow-rolls even after the expansion
rate of the universe becomes comparable to $m_\phi$.  Then, the
universe enters into the de Sitter phase again.  In this case, $\phi$
plays the role of the inflaton and the resultant cosmic density
perturbation becomes the same as the standard adiabatic result (if the
scale of the secondary inflation is properly chosen).\footnote
{If the $e$-folding number due to the secondary inflation is small,
however, this is not the case and the cosmic density fluctuation for
large scale is generated by the first inflation.  We do not consider
such a case in this paper.}
Thus, we assume that the initial amplitude of $\phi$ is smaller than
$M_*$.  In particular, we assume that $\phi_{\rm init}$ is small
enough so that the universe has a radiation-dominated epoch before
$\phi$ dominates the universe.  (This is the case if $\phi_{\rm init}$
is substantially smaller than $M_*$.)

\begin{figure}
    \begin{center}
        \scalebox{0.9}{\includegraphics{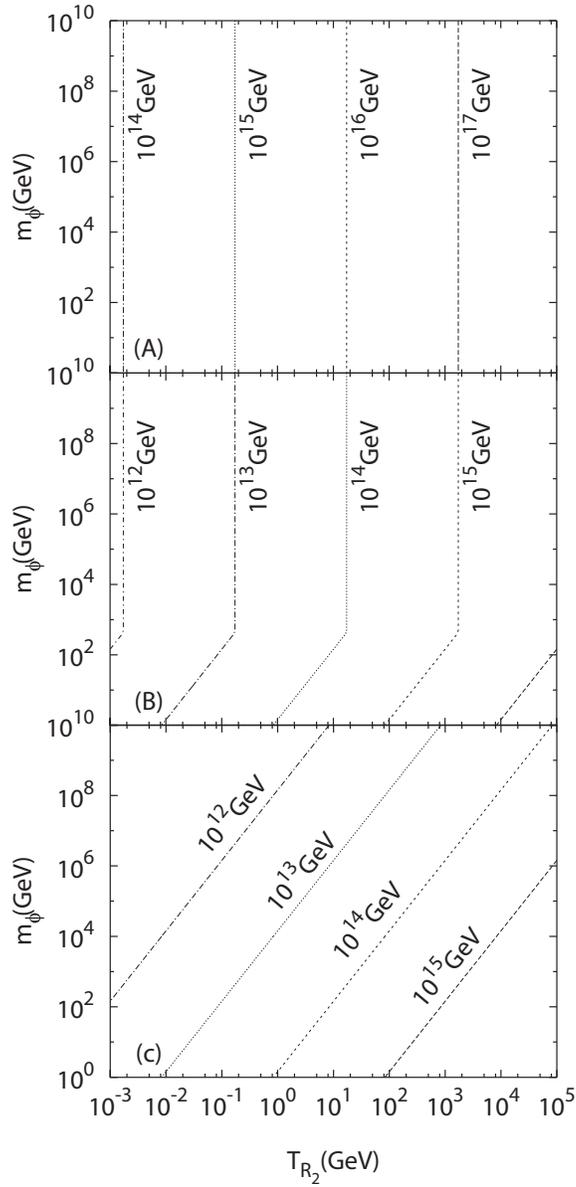}}
        \caption{Contours of minimum values of the initial amplitude 
        of the $\phi$ field with which the universe is once dominated
        by the energy density of $\phi$.  The vertical axis is the
        mass of $\phi$ and the horizontal axis is $T_{\rm R2}$.
        $T_{\rm R1}$ is taken to be (A) $T_{\rm R1}=1 \times
        10^{6}$GeV, (B) $T_{\rm R1}=1 \times 10^{10}$GeV and (C)
        $T_{\rm R1}=1 \times 10^{14}$GeV.}
        \label{fig:phimin}
    \end{center}
\end{figure} 

In Fig.\ \ref{fig:phimin}, we summarize the above results; on the
$T_{\rm R2}$ vs.\ $m_\phi$ plane, we plot the contours of the minimum
value of $\phi_{\rm init}$ with which the universe experiences the
$\phi$-dominated epoch.  In the following discussion, we assume that
the initial amplitude of the scalar field is large enough so that
there exists the $\phi$-dominated epoch.

Since the universe has to follow the standard big-bang scenario in
particular for the big-bang nucleosynthesis, however, the universe has
to become radiation-dominated after the $\phi$-dominated epoch.  This
is realized by the reheating due to the decay of $\phi$.  When $H\sim
\Gamma_\phi$, the $\phi$ field decays and its energy density is
converted to that of radiation.  The reheating temperature at the time
of the $\phi$-decay depends on the decay rate $\Gamma_\phi$.
Importantly, the reheating temperature has to be higher than $\sim 1\
{\rm MeV}$ so that the success of the standard big-bang
nucleosynthesis is not altered; we assume that $\Gamma_\phi$ is large
enough in order not to affect the standard nucleosynthesis scenario.

\begin{figure}[t]
    \begin{center}
        \scalebox{0.5}{\includegraphics{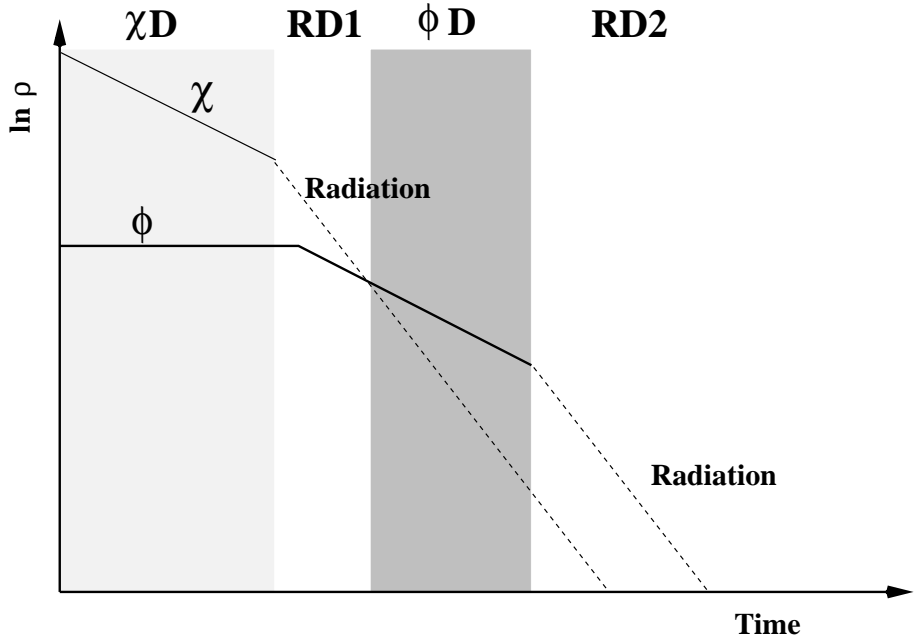}}
        \caption{Schematic picture of the thermal history of 
        the universe; evolution of the energy density of various
        components.  Here, we assumed that the $\phi$ field starts to
        oscillate in the RD1 epoch, but it may happen in the $\chi$D
        epoch.  The solid lines are for scalar condensations and the
        dotted ones are for radiation.  }
        \label{fig:history}
    \end{center}
\end{figure}

In summary, the thermal history is as follows:
\begin{itemize}
\item[1.] In the early stage, inflation occurs.\footnote
    {For the pre-big-bang \cite{PreBigBang} and the ekpyrotic
    \cite{ekpyrotic} scenarios, this may not be the case.  In these
    cases, we assume that some other mechanism generates the
    primordial fluctuation of $\phi$.}
    During inflation, the scalar field $\phi$ has a substantial
    initial amplitude so that it dominates the universe at the later
    stage.
\item[2.] After the inflation, the universe is dominated by the
    oscillating inflaton field.  (We call this epoch as ``$\chi$D''
    epoch.)  At some point, the inflaton decays and reheats the
    universe.
\item[3.] When the expansion rate of the universe becomes comparable
    to $m_\phi$, the $\phi$ field starts to oscillate.  This may
    happen during the $\chi$D or the radiation-dominated epoch.  We
    assume that the initial amplitude of $\phi$ is smaller than $\sim
    M_*$ so that there is no inflation induced by the $\phi$ field.
    In addition, if $\phi_{\rm init}$ is substantially smaller than
    $M_*$, the universe experiences a radiation-dominated epoch before
    $\phi$ dominates the universe.  (We call this epoch as ``RD1''
    epoch.)
\item[4.] As the universe expands, $\phi$ eventually dominates the
    universe since $\rho_\phi\propto a^{-3}$ while $\rho_{\rm
    rad}\propto a^{-4}$ (where $\rho_{\rm rad}$ is the energy density
    of radiation).  We call this epoch ``$\phi$D'' epoch.
\item[5.] When $H\sim\Gamma_\phi$, the $\phi$ field decays and reheats
    the universe.  Then the radiation-dominated universe is realized
    again.  (We call this epoch ``RD2'' epoch.)  After this epoch, the
    evolution of the universe is the same as the standard big-bang
    case as far as the cosmic density perturbations are not concerned.
\end{itemize}
The scenario is schematically shown in Fig.\ \ref{fig:history}.  (In
the following, the subscripts ``$\chi$D,'' ``RD1,'' ``$\phi$D,'' and
``RD2'' will be used for variables in each epochs.)

\section{Evolutions of the Perturbations}
\label{sec:evolution}
\setcounter{equation}{0}

In the previous section, we summarized the scenario and discussed the
evolutions of the background variables (i.e., the zero-modes).  Our
purpose is, however, to understand the effects of the fluctuation of
the amplitude of the scalar field $\phi$ on the cosmic density
perturbations.  Thus, in this section, we discuss the evolutions of
the perturbations in detail.  Importantly, properties of the density
perturbations are determined at the epoch much earlier than the
electron decoupling (i.e., $T\sim m_e$, with $m_e$ being the electron
mass) and the neutrino freeze-out.  Thus, in this section, we study
the behaviors of the perturbations in such an epoch.  In this paper,
we give the formalism in the conformal Newtonian gauge.  (We follow
the notation and convention of \cite{hu_PhD_thesis}.)

In discussing the evolutions of the perturbations, it is convenient to
define the variable $\delta_X$ as
\begin{eqnarray}
\delta_X \equiv \delta \rho_X / \rho_X,
\end{eqnarray}
where the subscript $X$ denotes the individual components (like
radiation, CDM, baryon, and so on) and $\delta\rho_X$ is the density
perturbation of the component $X$.  In addition, in the conformal
Newtonian gauge, the perturbed line element is given by
\begin{eqnarray}
    ds^2 &=& - (1 + 2\Psi) dt^2
    + \left( \frac{a}{a_0} \right)^2 (1 + 2\Phi) \delta_{ij} dx^i dx^j 
    \nonumber \\
     &=& 
    \left( \frac{a}{a_0} \right)^2 
    \left[ - (1 + 2\Psi) d\tau^2 + (1 + 2\Phi) 
    \delta_{ij} dx^i dx^j \right],
    \label{ds}
\end{eqnarray}
where $\tau$ is the conformal time, $a_0$ is the scale factor at the
present time, and $\Psi$ and $\Phi$ are the metric perturbations.

Substituting Eq.\ (\ref{ds}) into the Einstein equation, we obtain the
generalized Poisson equation for $\Phi$:
\begin{eqnarray}
    k^2 \Phi 
    = \frac{1}{2M_*^2} \left( \frac{a}{a_0} \right)^2  
    \rho_{\rm tot} \left[ 
        \delta_{\rm tot} + \frac{3 \mathcal{H}}{k} 
        (1+\omega_{\rm tot}) V_{\rm tot}
    \right],  
    \label{poisson}
\end{eqnarray}
and other equation relating $\Phi$, $\Psi$, and the anisotropic stress
perturbation of the total matter $\Pi_{\rm tot}$:
\begin{eqnarray}
    k^2 (\Psi + \Phi) = - \frac{1}{M_*^2} 
    \left( \frac{a}{a_0} \right)^2  
    p_{\rm tot} \Pi_{\rm tot}.
    \label{Psi+Phi=0}
\end{eqnarray}
Here, ``tot'' denotes the total matter and the variable $V_X$ denotes
the velocity perturbation of the component $X$.  In addition, $k$ is
the comoving momentum, $\omega_{\rm tot}\equiv\rho_{\rm tot}/p_{\rm
tot}$ the equation-of-state parameter for the total matter, and
\begin{eqnarray}
    \mathcal{H} \equiv \frac{1}{a} \frac{da}{d\tau}.
\end{eqnarray}

As indicated in the above equations, evolutions of the metric
perturbations are related to those of the density, velocity, and other
perturbations of various components.  Here, we consider the universe
with temperature higher than $T\simeq m_e$.  In such a situation,
various charged particles become relativistic and are thermally
produced.  Then, all the relativistic components (like photon,
electron, quarks, neutrinos, and so on) can be regarded as a single
fluid with very short mean free path, much shorter than the horizon
scale.  We call such a component as ``radiation'' and use the
subscript ``$r$'' for it.

When the mean free path of the radiation is very short, perturbation
of the radiation becomes locally isotropic and the anisotropic stress
perturbation vanishes:
\begin{eqnarray}
    \Pi_r= 0.
\end{eqnarray}
Then, the equations for the density and velocity perturbations of the
radiation are given by
\begin{eqnarray}
    \delta_r' &=& -\frac{4}{3} k V_r - 4 \Phi', 
    \label{dr'} \\
    V_r' &=& \frac{1}{4} k \delta_r + k\Psi,
    \label{Vr'}
\end{eqnarray}
where the ``prime'' denotes the derivative with respect to the
conformal time $\tau$.  In addition, if a very weakly interacting
non-relativistic component exists, its perturbations obey the
following equations:
\begin{eqnarray}
    \delta_m' &=& -k V_m - 3\Phi',
    \label{eq:cdm} \\
    V_m'  &=& - \mathcal{H} V_m + k \Psi,
    \label{eq:cdm_velocity}
\end{eqnarray}
where the subscript ``$m$'' is for non-relativistic matters.  Notice
that, when a scalar field is oscillating, the equation-of-state
parameter of the scalar condensation vanishes and hence the density
and velocity perturbations of the scalar field also obey Eqs.\ 
(\ref{eq:cdm}) and (\ref{eq:cdm_velocity}).

In our scenario, there are several possible sources of the density
perturbations.  The first one is, as in the conventional case,
fluctuation of the inflaton amplitude.  In our case, however,
amplitude of $\phi$ may also fluctuate and such a fluctuation becomes
second source of the density fluctuations.  In addition, if there
exists independent scalar field (called $\psi$) from which the baryon
or the CDM is generated, we also have to take account of its effects.
(In the following, we consider the cases with and without this scalar
field $\psi$.)  In the following, primordial fluctuations of these
scalar fields are assumed to be uncorrelated.  In the framework of the
linear perturbation theory, evolutions of the perturbations from the
primordial fluctuations of these scalar fields are independent and
hence we can study their evolutions separately.  Evolutions of the
perturbations from the inflaton fluctuation are intensively studied;
such a fluctuation provides adiabatic density fluctuations.  In
addition, density perturbation induced from the fluctuation of $\psi$
is also well known as far as $\psi$ is a minor component in the RD2
epoch; primordial fluctuation of $\psi$ induces (purely) isocurvature
fluctuation to the baryonic or the CDM component.

Thus, in the following, we study the evolutions of the density
perturbations induced by the fluctuation of the amplitude of $\phi$.
For this purpose, we decompose the scalar field as
\begin{eqnarray}
    \phi (t,k) = \phi (t) + \delta\phi (k,t),
\end{eqnarray}
where $\delta\phi$ denotes the fluctuation of $\phi$.  (We use the
same notation for the total amplitude and for the zero-mode because we
believe there is no confusion.)  In particular we study evolutions in
very early universe in order to understand how the fluctuations behave
when the universe enters into the RD2 epoch.  In realistic scenarios,
the scale we are interested in (i.e., the scale corresponding to the
multipole $l\lesssim O(1000)$) is far outside of the horizon at this
epoch and hence we can expand the solutions to the equations as
functions of $k\tau$.  In addition, combining the relation $\Pi_r=0$
with Eq.\ (\ref{Psi+Phi=0}), we obtain $\Phi =-\Psi$.  In the
following, we use this relation to eliminate $\Phi$.

In order to discuss the evolutions of the fluctuations, it is
important to clarify the equation-of-state parameters of the
individual components in the universe.  If all the components behave
as the relativistic or the non-relativistic matter, evolutions of the
perturbations are described by Eqs.\ (\ref{dr'}) $-$
(\ref{eq:cdm_velocity}).  In this case, it is convenient to
distinguish the photon (or any other components) from the decay
product of $\phi$ from that from the inflaton field, which we call
$\gamma_\phi$ and $\gamma_\chi$, respectively.\footnote
{In fact, these photons are mixed each other and they cannot be
defined separately.  In other words, their velocity perturbations
should be the same since they form a single fluid.  Even so, the
following arguments are unchanged as far as we consider the leading
terms in the density perturbations since the velocity perturbation is
at most $O(k\tau)$.  In the following discussion, $\gamma_\phi$ and
$\gamma_\chi$ should be understood as representatives of the
components which are and are not generated from the decay product of
$\phi$, respectively.}
In order to consider $\delta_{\gamma_\phi}$ in the RD2 epoch, we can
neglect $\gamma_\chi$ since the CMB radiation at this epoch is
dominantly from the $\phi$ field.  Then, we find that, in the RD2
epoch, $\Psi$ and $\delta_{\gamma_\phi}$ become constant while
$V_{\gamma_\phi}=O(k\tau)$ up to higher order corrections.  Indeed,
combining Eq.\ (\ref{poisson}) with Eqs.\ (\ref{dr'}) and (\ref{Vr'}),
and using $\delta_{\rm tot}=\delta_{\gamma_\phi}$ and $V_{\rm
tot}=V_{\gamma_\phi}$, we obtain
$V_{\gamma_\phi}=-\frac{1}{2}k\tau\Psi_{\rm RD2}^{(\delta\phi)}$ and
\begin{eqnarray}
    \delta_{\gamma_\phi} = -2 \Psi_{\rm RD2}^{(\delta\phi)},
\end{eqnarray}
where $\Psi_{\rm RD2}^{(\delta\phi)}$ is the metric perturbation
induced by the primordial fluctuation of the amplitude of $\phi$.  (In
the following, the superscript ``$(\delta\phi)$'' is for perturbations
generated from the primordial fluctuation of $\phi$.)  As we
mentioned, $\Psi_{\rm RD2}^{(\delta\phi)}$ is constant up to a
correction of $O(k^2\tau^2)$.

Behavior of $\delta_{\gamma_\chi}$ is also easily understood.  In
discussing the effects of the primordial fluctuation of $\phi$, we
neglect the initial fluctuation of the inflaton field and hence
$\delta_{\gamma_\chi}\rightarrow 0$ in the deep RD1 epoch.  In
addition, from Eqs.\ (\ref{dr'}) and (\ref{Vr'}), $V_{\gamma_\chi}$
becomes higher order in $k\tau$ than $\delta_{\gamma_\chi}$ and
$\Psi$.  Thus, we obtain
\begin{eqnarray}
    \delta_{\gamma_\chi} = 4 \Psi^{(\delta\phi)}.
\end{eqnarray}
The above relation holds in the RD1, $\phi$D, and RD2 epochs up to
corrections of $O(k^2\tau^2)$.

For the following discussion, it is convenient to define the entropy
perturbation between $\gamma_\phi$ and $\gamma_\chi$:
\begin{eqnarray}
    S_{\phi\chi}^{(\delta\phi)} 
    \equiv \frac{3}{4} \delta_{\gamma_\phi}
    - \frac{3}{4} \delta_{\gamma_\chi}
    = -\frac{9}{2} \Psi_{\rm RD2}^{(\delta\phi)}.
\end{eqnarray}
For superhorizon modes, this entropy perturbation is a constant of
time \cite{hu_PhD_thesis}.  In addition, for example for the case
where the $\phi$ field starts to oscillate during the $\chi$D epoch,
\begin{eqnarray}
    S_{\phi\chi}^{(\delta\phi)} = 
    \left[ \frac{3}{4} \delta_{\gamma_\phi}
        - \frac{3}{4} \delta_{\gamma_\chi} \right]_{\rm RD2}
    = \left[ \delta_{\phi}
        - \frac{3}{4} \delta_{\gamma_\chi} \right]_{\rm \phi D, RD1}
    = \left[ \delta_{\phi}
        - \delta_{\chi} \right]_{\rm \chi D},
\end{eqnarray}
where the last equality holds for the epoch when $\phi$ and the
inflaton field are both oscillating.

Density (and other) fluctuations in the RD2 epoch are generally
parameterized by using $S_{\phi\chi}^{(\delta\phi)}$.  If a component
$X$ is generated from the decay product of $\phi$, then there is no
entropy between the photon (i.e., $\gamma_\phi$) and $X$.  On the
contrary, if some other scalar field $\psi$ generates $X$, the entropy
between the photon and $X$ is the same as
$S_{\phi\chi}^{(\delta\phi)}$.  Thus, if all the components in the
universe are generated from $\phi$, the density fluctuations become
purely adiabatic and
\begin{eqnarray}
    \left[ \delta_\gamma^{(\delta\phi)} \right]_{\rm RD2} 
    = \frac{4}{3}\left[ \delta_b^{(\delta\phi)} \right]_{\rm RD2}
    = \frac{4}{3}\left[ \delta_c^{(\delta\phi)} \right]_{\rm RD2}
    = -2 \Psi_{\rm RD2}^{(\delta\phi)},
\end{eqnarray}
where the subscripts $\gamma$, $b$, and $c$ are for the photon,
baryon, and CDM, respectively.  In this case, the isocurvature
perturbation in the $\phi$ field is converted to the purely adiabatic
density perturbation after the decay of $\phi$
\cite{PRD42-313,NPB626-395,PLB524-5,PLB522-215}.  On the contrary, if
the baryon asymmetry is generated by the scalar field $\psi$, the
entropy between the radiation and the baryon becomes
$S_{\phi\chi}^{(\delta\phi)}$ and hence \cite{PLB522-215}
\begin{eqnarray}
    \left[ \delta_\gamma^{(\delta\phi)} \right]_{\rm RD2} 
    = \frac{4}{3}\left[ \delta_c^{(\delta\phi)} \right]_{\rm RD2}
    = -2 \Psi_{\rm RD2}^{(\delta\phi)},~~~
    \left[ \delta_b^{(\delta\phi)} \right]_{\rm RD2}
    = \frac{3}{4} \left[ \delta_\gamma^{(\delta\phi)} \right]_{\rm RD2} 
    + \frac{9}{2} \Psi_{\rm RD2}^{(\delta\phi)},
    \label{Sb}
\end{eqnarray}
and in the case where $\psi$ is responsible for the CDM while the
baryon number is somehow generated from the decay product of $\phi$,
\begin{eqnarray}
    \left[ \delta_\gamma^{(\delta\phi)} \right]_{\rm RD2} 
    = \frac{4}{3}\left[ \delta_b^{(\delta\phi)} \right]_{\rm RD2}
    = -2 \Psi_{\rm RD2}^{(\delta\phi)},~~~
    \left[ \delta_c^{(\delta\phi)} \right]_{\rm RD2}
    = \frac{3}{4} \left[ \delta_\gamma^{(\delta\phi)} \right]_{\rm RD2} 
    + \frac{9}{2} \Psi_{\rm RD2}^{(\delta\phi)}.
    \label{Sc}
\end{eqnarray}
In addition, if the baryon and the CDM are both generated from sources
other than $\phi$, we obtain
\begin{eqnarray}
    \left[ \delta_\gamma^{(\delta\phi)} \right]_{\rm RD2} 
    = -2 \Psi_{\rm RD2}^{(\delta\phi)},~~~
    \left[ \delta_b^{(\delta\phi)} \right]_{\rm RD2}
    = \left[ \delta_c^{(\delta\phi)} \right]_{\rm RD2}
    = \frac{3}{4} \left[ \delta_\gamma^{(\delta\phi)} \right]_{\rm RD2} 
    + \frac{9}{2} \Psi_{\rm RD2}^{(\delta\phi)}.
    \label{Sbc}
\end{eqnarray}
It is important to notice that, for the cases given in Eqs.\ 
(\ref{Sb}) $-$ (\ref{Sbc}), the isocurvature perturbation is
correlated with the adiabatic perturbation.

In the case with the $\psi$ field which is responsible for the baryon
or the CDM, more careful considerations are needed.  Adopting the
parabolic potential for $\psi$, the fluctuation of the $\psi$ field
obeys
\begin{eqnarray}
    \ddot{\delta\psi} + 3H \dot{\delta\psi}
    + \left[k^2 \left(\frac{a}{a_0}\right)^2
        + m_\psi^2 \right] \delta\psi 
    = -2 m_\psi^2 \psi  \Psi
    + \dot{\psi} ( \dot{\Psi} - 3 \dot{\Phi} ).
    \label{eom_psi}
\end{eqnarray}
Solving this equation, we see that the fluctuation of $\psi$ at the
time of $H\sim m_\psi$ is of the order of $[\psi\Psi]_{H\sim m_\psi}$.
If the $\psi$ field starts to oscillate in the RD1 epoch, energy
density of $\psi$ behaves as that of non-relativistic components even
when the metric perturbation $\Psi$ is negligibly small.  In this
case, $[\psi\Psi]_{H\sim m_\psi}$ is small enough to be neglected and
the former discussion applies.  If $\psi$ starts to oscillate in the
$\phi$D epoch, on the contrary, the situation changes.  In this case,
the $\psi$ field is slowly rolling when the $\phi$D epoch starts and
$\delta\psi$ at the time of the $\phi$-domination is negligibly small.
Then, solving Eq.\ (\ref{eom_psi}) for the superhorizon mode, we
obtain
\begin{eqnarray}
     \delta\psi^{(\delta\phi)}_{\rm \phi D} = 
     \dot{\psi} \Psi^{(\delta\phi)}_{\rm \phi D} t,
     \label{delpsi}
\end{eqnarray}
where we used the fact that $\dot{\Psi}$ and $\dot{\Phi}$ are small
enough to be neglected in the $\phi$D epoch.  With Eq.\ 
(\ref{delpsi}), we can calculate the density fluctuation of $\psi$ as
$[\rho_\psi\delta_\psi]_{\rm \phi
D}=-3H\dot{\psi}^2\Psi^{(\delta\phi)}_{\rm \phi D} t$.  Importantly,
the relation (\ref{delpsi}) holds even after the $\psi$ field starts
to oscillate, as far as $H$ is substantially larger than the decay
rate of $\psi$.  So, we can calculate the entropy between $\psi$ and
$\phi$ in the $\phi$D epoch after the $\psi$ field starts to
oscillate; using the relation $[\delta\phi]_{\rm\phi D} = -2
\Psi^{(\delta\phi)}_{\rm \phi D}$ and
$\langle\dot{\psi}^2\rangle=\rho_\psi$, we see that the entropy
between $\psi$ and $\phi$ vanishes.  Thus, if the $\psi$ field is
slow-rolling when the $\phi$D epoch starts, density perturbation of
the component generated from $\psi$ becomes adiabatic.  If the $\psi$
field starts to oscillate just at the time when the $\phi$D epoch
starts, the correlated mixture of the adiabatic and isocurvature
perturbations is generated but the ratio of the metric perturbation to
the entropy perturbation may deviate from that given in Eqs.\ 
(\ref{Sb}) $-$ (\ref{Sbc}).

Before closing this section, let us estimate the size of
$S_{\phi\chi}^{(\delta\phi)}$.  As we mentioned before, primordial
fluctuation of the $\phi$ field is probably generated during the
inflation.  In this case, calculating the two-point function of $\phi$
in the de Sitter background, we obtain
\begin{eqnarray}
    \langle 0 | \delta\phi (t, \vec{x}) \delta\phi (t, \vec{y}) 
    | 0 \rangle_{\rm inf} =
    \int \frac{d k}{k}
    |\delta\phi (t, \vec{k})|^2
    e^{i \vec{k} (\vec{x}-\vec{y})},
\end{eqnarray}
where the Fourier amplitude for the case of $m_\phi\ll H_{\rm inf}$ is
given by
\begin{eqnarray}
    \delta\phi (t, \vec{k}) = 
    \left( \frac{k}{2aH_{\rm inf}}
    \right)^{2m_\phi^2/3H_{\rm inf}^2}
    \left[ \frac{H_{\rm inf}}{2\pi} \right]_{k=aH_{\rm inf}}.
\end{eqnarray}
(The subscript ``inf'' means that the correlation function is evaluated
during the inflation.)  Once the wavelength becomes longer than the
horizon scale, the scalar-field fluctuation obeys the following
equation:
\begin{eqnarray}
    \ddot{\delta\phi} + 3H \dot{\delta\phi}
    + \left[k^2 \left(\frac{a}{a_0}\right)^2
        + m_\phi^2 \right] \delta\phi 
    = -2 m_\phi^2 \phi \Psi 
    + \dot{\phi} (\dot{\Psi} - 3 \dot{\Phi}).
    \label{dphi_eom}
\end{eqnarray}

First, let us consider the case where $\phi_{\rm init}\ll M_*$.  In
this case, when the $\phi$ field starts to move, the energy density of
$\phi$ is much smaller than the total energy density of the universe.
In studying the density perturbations generated by the primordial
fluctuations of $\phi$, the metric perturbations are negligibly small
until the energy density of $\phi$ becomes comparable to the total
energy density of the universe.  In this case, we can neglect the
terms containing the metric perturbations in Eq.\ (\ref{dphi_eom}).
Furthermore, the $k^2$-term is also irrelevant for the superhorizon
perturbations.  Then, we can see that, with the above approximations,
$\delta\phi$ obeys the same equation as the zero-mode.  Neglecting the
decaying mode, we obtain the relation
\begin{eqnarray}
    \delta\phi(t, k) = 
    \frac{\delta\phi_{\rm init}(k)}{\phi_{\rm init}} \phi(t),
\end{eqnarray}
where $\delta\phi_{\rm init}$ is the initial value of the fluctuation
of $\phi$, which is $H_{\rm inf}/2\pi$.  In addition,
$\delta_\chi\rightarrow 0$ (and hence $\delta_{\gamma_\chi}\rightarrow
0$) in the early universe, and hence the initial value of
$S_{\phi\chi}^{(\delta\phi)}$ is estimated as
\begin{eqnarray}
    S_{\phi\chi}^{(\delta\phi)} (k) = 
    \frac{\delta\rho_\phi (t, k)}{\rho_\phi (t)} - \delta_\chi (t)
    = \frac{2\delta\phi_{\rm init} (k)}{\phi_{\rm init}}.
    \label{s_pc}
\end{eqnarray}
Thus, the entropy $S_{\phi\chi}^{(\delta\phi)}$ increases as the
initial amplitude of $\phi$ decreases.  If the initial amplitude of
$\phi$ becomes close to $M_*$, the energy density of $\phi$ becomes
comparable to the total energy density of the universe.  Then, the
metric perturbation $\Psi$ becomes comparable to the density
perturbation of the $\phi$ field $\delta_\phi$.  In this case, sizable
$\Psi$ generates the density perturbations of other components and
hence the relation (\ref{s_pc}) may not hold.  Even in this case,
however, Eq.\ (\ref{s_pc}) provides a good order-of-estimation of the
primordial entropy perturbation between $\phi$ and $\chi$ at the time
when the $\phi$ field starts to oscillate; as far as there is no
inflation caused by $\phi$, $S_{\phi\chi}^{(\delta\phi)}\sim
O(\delta\phi_{\rm init}/M_*)$ when $\phi_{\rm init}\sim M_*$.

\section{CMB Angular Power Spectrum: Overview}
\label{sec:cl}
\setcounter{equation}{0}

In this section, we study the CMB anisotropy generated from
fluctuations of various scalar fields.  The CMB anisotropy is
characterized by the angular power spectrum $C_l$ which is defined as
\begin{eqnarray}
    \left\langle \Delta T(\vec{x}, \vec{\gamma}) 
        \Delta T(\vec{x}, \vec{\gamma}')  \right\rangle_{\vec{x}}
    = \frac{1}{4\pi} 
    \sum_l (2l+1) C_l P_l (\vec{\gamma} \cdot \vec{\gamma}'),
\end{eqnarray}
with $\Delta T (\vec{x}, \vec{\gamma})$ being the temperature
fluctuation of the CMB radiation pointing to the direction
$\vec{\gamma}$ at the position $\vec{x}$ and $P_l$ is the Legendre
polynomial.  The angular power spectrum $C_l$ depends on the
properties of the primordial density fluctuations and hence it
provides probes for studying how the density perturbations are
generated in the early universe.

In the scenario we consider, there are three possible sources of the
density perturbations; the primordial fluctuations of the scalar
fields $\chi$, $\phi$, and $\psi$.  Since there is no correlation
among these fields, the CMB anisotropies from these fluctuations are
uncorrelated and the resultant CMB power spectrum is given in the form
\begin{eqnarray}
    C_l 
    = C_l^{(\delta\chi)} + C_l^{(\delta\phi)} + C_l^{(\delta\psi)},
\end{eqnarray}
where $C_l^{(\delta \chi)}$, $C_l^{(\delta \phi)}$ and
$C_l^{(\delta\psi)}$ are contributions from the primordial
fluctuations of the inflaton field $\chi$, the $\phi$ field and the
$\psi$ field, respectively.\footnote
{If all the components in the universe are generated from the decay
product of $\phi$, the $\psi$ field is irrelevant and
$C_l^{(\delta\psi)}$ does not exist.}

The inflaton contribution $C_l^{(\delta\chi)}$ is the same as the
adiabatic result while $C_l^{(\delta\psi)}$ becomes the purely
isocurvature one.  Thus, in the following, we study the properties of
$C_l^{(\delta\phi)}$.

Effects of the primordial fluctuation of the $\phi$ field depend on
the scenario as we discussed.  If all the components in the universe
originate to $\phi$, then the density perturbations become adiabatic.
In this case, it is difficult to experimentally distinguish
$C_l^{(\delta\phi)}$ from the adiabatic CMB power spectrum generated
from the inflaton fluctuations.  Even in this case, however, there are
important implications in studying scenarios of generating the cosmic
density fluctuations, like inflation, as will be discussed in Section
\ref{sec:noentropy}.

On the contrary, if the scalar field $\psi$ exists which is
responsible for the baryon or the CDM, $C_l^{(\delta\phi)}$ may become
different from the angular power spectrum from the purely adiabatic
density perturbation since the adiabatic and isocurvature
perturbations can be generated with cross-correlation.  Although we
have considered the entropy perturbations in the baryonic and the CDM
sectors separately in Eqs.\ (\ref{Sb}) $-$ (\ref{Sbc}), we found that
the resultant $C_l^{(\delta\phi)}$ is primarily determined by the
entropy between the non-relativistic matter and radiation; if we
parameterize the density perturbation of the non-relativistic
component as
\begin{eqnarray}
    \left[ \delta_m^{(\delta\phi)} \right]_{\rm RD2} \equiv
    \left[ (\Omega_b/\Omega_m) \delta_b^{(\delta\phi)} 
        + (\Omega_c/\Omega_m) \delta_c^{(\delta\phi)} 
    \right]_{\rm RD2} =
    \frac{3}{4} \left[ \delta_\gamma^{(\delta\phi)} \right]_{\rm RD2} 
    + \kappa_m \Psi_{\rm RD2}^{(\delta\phi)},
\end{eqnarray}
where $\Omega_b$, $\Omega_c$ and $\Omega_m$ are the (present) density
parameters for the baryon, the CDM, and the non-relativistic component
(and hence $\Omega_m=\Omega_b+\Omega_c$), the parameter $\kappa_m$
determines the shape of $C_l^{(\delta\phi)}$.  With the relations
(\ref{Sb}), (\ref{Sc}), and (\ref{Sbc}), $\kappa_m$ becomes
$\frac{9}{2}(\Omega_b/\Omega_m)$, $\frac{9}{2}(\Omega_c/\Omega_m)$,
and $\frac{9}{2}$, respectively. On the contrary, if the $\psi$ field
is slow-rolling when the $\phi$D epoch starts, $\kappa_m$ vanishes.
In addition, if $\psi$ accidentally starts to oscillate just at the
beginning of the $\phi$D epoch, $\kappa_m$ may take any value between
0 and $\frac{9}{2}$.  This may happen, however, when the expansion
rate at the beginning of the $\phi$D epoch accidentally becomes
comparable to $m_\psi$.

\begin{figure}[t]
    \begin{center}
        \scalebox{1.0}{\includegraphics{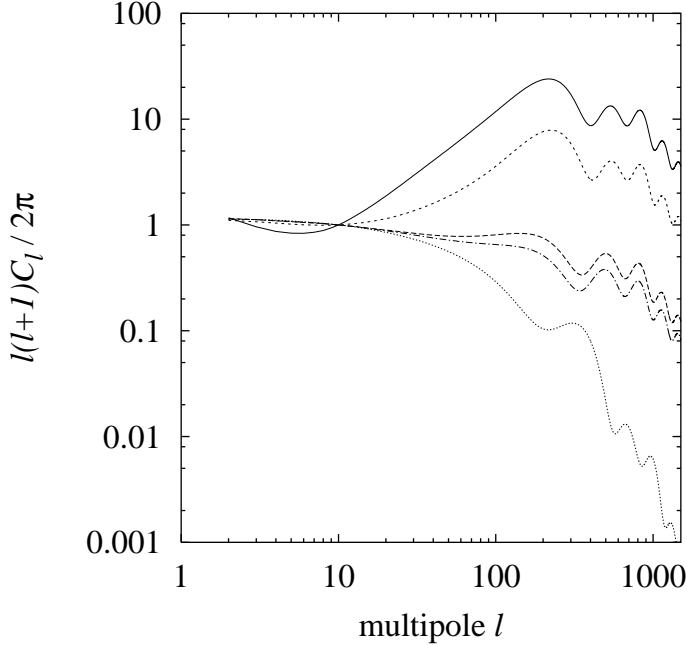}}
        \caption{The angular power spectrum with correlated mixture 
        of the adiabatic and isocurvature perturbations in the
        baryonic sector (solid line), in the CDM sector (long-dashed
        line), and in the baryonic and CDM sectors (dot-dashed line).
        (See Eqs.\ (\ref{Sb}), (\ref{Sc}), and (\ref{Sbc}),
        respectively.)  We also show the CMB angular power spectrum in
        the purely adiabatic (short-dashed line) and isocurvature
        density perturbations (dotted line), i.e.,
        $C_l^{(\delta\chi)}$ and $C_l^{(\delta\psi)}$.  We consider
        the flat universe with $\Omega_bh^2=0.019$, $\Omega_m=0.3$,
        and $h=0.65$, and the initial power spectral indices for
        primordial density perturbations are all assumed to be 1
        (i.e., we adopt scale-invariant initial power spectra).  The
        overall normalizations are taken as
        $[l(l+1)C_l/2\pi]_{l=10}=1$.}
        \label{fig:Cl's}
    \end{center}
\end{figure}

In Fig.\ \ref{fig:Cl's}, we plot the angular power spectrum with the
correlated mixture of the adiabatic and isocurvature perturbations in
the baryonic and/or CDM sector, i.e., the cases with the relations
given in Eqs.\ (\ref{Sb}) $-$ (\ref{Sbc}).  Here, we consider the flat
universe with $\Omega_bh^2=0.019$, $\Omega_m=0.3$, and $h=0.65$
\cite{primack}, where $h$ is the Hubble constant in units of 100
km/sec/Mpc.  For comparison, we also plot the angular power spectrum
for the purely adiabatic and isocurvature cases (i.e.,
$C_l^{(\delta\chi)}$ and $C_l^{(\delta\psi)}$).  As one can see, the
CMB angular power spectrum strongly depends on properties of the
primordial density perturbations.  If there exists correlated entropy
between the baryon and other components with the relation (\ref{Sb}),
negative interference between the adiabatic and isocurvature;
perturbations suppresses $C_l$ at lower multipole while the effect of
the isocurvature perturbation becomes too small to affect the
structure at high multipole.  As a result, the angular power spectrum
is enhanced at the high multipole rather than at the low multipole.
If the effect of the entropy perturbation becomes more efficient, then
$C_l^{(\delta\psi)}$ at high multipole is suppressed relative to that
at low multipole like in the purely isocurvature case.  This happens
when the entropy perturbation is in the CDM component with the
condition given in Eq.\ (\ref{Sc}).  In addition, with the relation
given in Eq.\ (\ref{Sbc}), the correlated entropy becomes more
effective than the case where only the CDM sector has the correlated
entropy.  Then, the acoustic peaks are more suppressed relative to the
Sachs-Wolfe (SW) tail.

\begin{figure}
    \begin{center}
        \scalebox{0.75}{\includegraphics{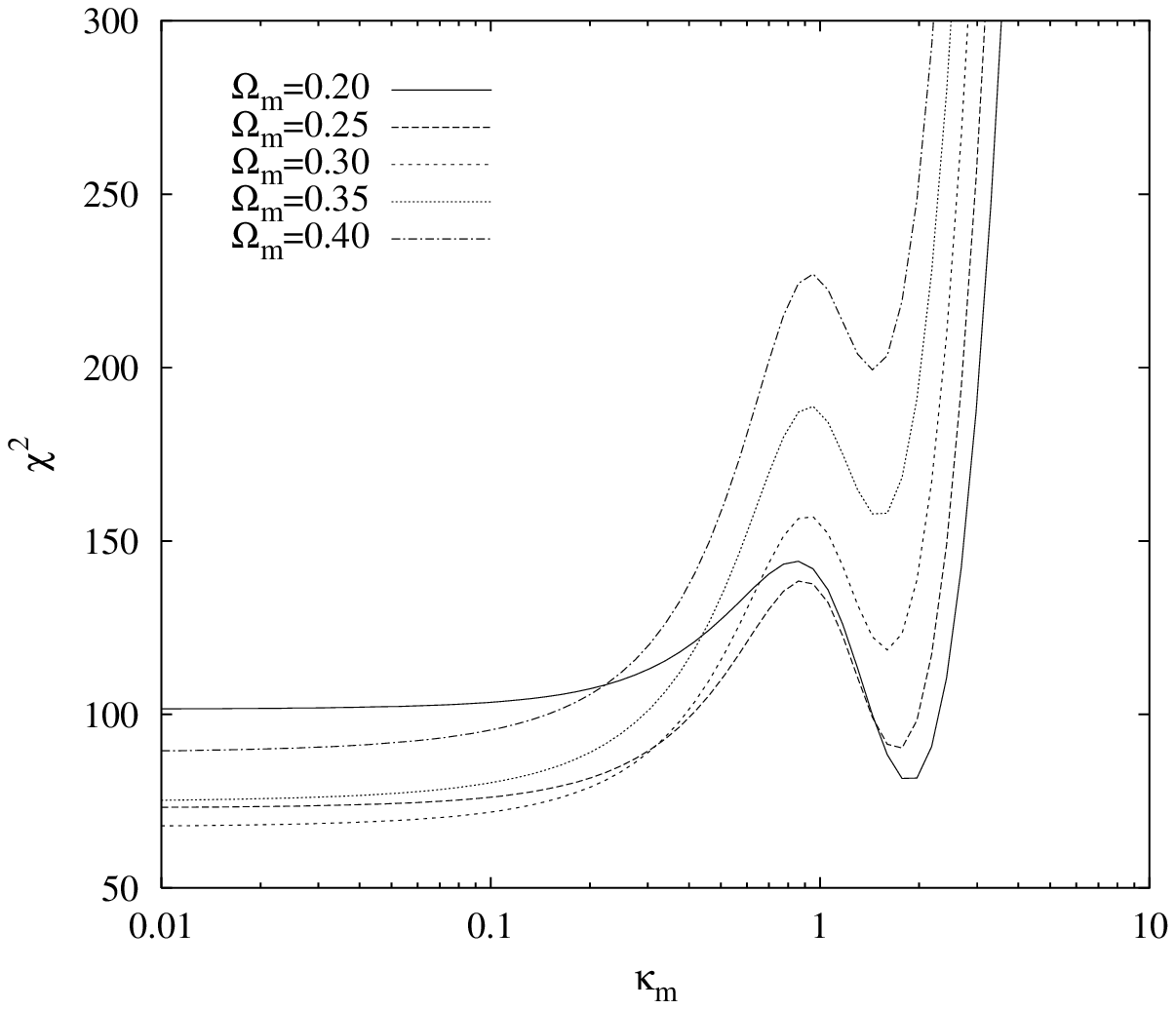}}
        \caption{$\chi^2$ as a function of $\kappa_m$.  We 
        take $\Omega_bh^2 = 0.019$, $h=0.65$, and the values of
        $\Omega_m$ are shown in the figure.  The flat universe is
        assumed.  Notice that the $\kappa_m$ parameter is smaller than
        $\frac{9}{2}$ in our scenario.}
        \label{fig:chi2kappa}
    \end{center}
\end{figure} 

As discussed in the previous section, in the general case, the
$\kappa_m$ parameter may vary between 0 and $\frac{9}{2}$, and hence
the CMB anisotropy generated from the primordial fluctuation of $\phi$
changes its behavior as $\kappa_m$ varies.  In particular, if
$\delta\phi_{\rm init}$ is the only source of the cosmic density
perturbations, too large $\kappa_m$ results in angular power spectrum
which is inconsistent with observations.  In this case, we can derive
a constraint on the $\kappa_m$ parameter.

In order to derive the constraint, we calculate the goodness-of-fit
parameter $\chi^2=-2\ln L$, where $L$ is the likelihood function, as a
function of $\kappa_m$.  In our calculation, the offset log-normal
approximation is used \cite{Bond_Jaffe_Knox_2000}.  We use a data set
consisting of 65 data points; 24 from COBE/DMR \cite{APJ464L1} in the
range $2 \leq l \leq 25$, 19 from BOOMERanG \cite{aph0104460} in the
range $76\leq l\leq 1025$, 13 from MAXIMA \cite{aph0104459} in the
range $36\leq l\leq 1235$, and 9 from DASI \cite{aph0104489} in the
range $104\leq l\leq 864$.  From the figure, we can read off the
constraint on $\kappa_m$. Requiring $\chi\leq 84$, which corresponds
to 95 \% C.L.\ allowed region for the $\chi^2$ statistics with 64
degrees of freedom, we obtain the constraint $\kappa_m\leq 0.3$ for
$\Omega_m=0.3$, and $1.7\leq \kappa_m\leq 2.0$ for $\Omega_m=0.2$.
Interestingly, the $\chi^2$ variable is relatively suppressed when
$\kappa_m\sim 2$.  This is because the heights of the acoustic peaks
relative to the SW tail once increase then decrease as we increase the
$\kappa_m$ parameter.  In particular, when $\kappa_m\sim 2$, the
heights of the acoustic peaks relative to the SW tail become similar
to those from the adiabatic density perturbations which well agrees
with the observations.  Of course, if the effects of
$C_l^{(\delta\chi)}$ or $C_l^{(\delta\psi)}$ become important, the
above constraint on $\kappa_m$ does not apply.

\section{Case without Entropy Perturbations}
\label{sec:noentropy}
\setcounter{equation}{0}

If all the component of the universe (i.e., the photon, baryon, CDM,
neutrino, and so on) are generated from the decay product of $\phi$,
there is no entropy between any of two components.  In this case, the
resultant angular power spectrum is given in the form
\begin{eqnarray}
C_l = C_l^{(\delta\chi)} + C_l^{(\delta\phi)}.
\end{eqnarray}
Importantly, the two contributions $C_l^{(\delta\chi)}$ and
$C_l^{(\delta\phi)}$ are both from adiabatic perturbations.  Thus, for
the case where scale dependence of the sources of $C_l^{(\delta\chi)}$
and $C_l^{(\delta\phi)}$, i.e., $\Psi^{(\delta\chi)}$ and
$S_{\phi\chi}^{(\delta\phi)}$, are the same,
$C_l^{(\delta\phi)}\propto C_l^{(\delta\chi)}$.  In this case, it is
impossible to test the scenario with $\phi$ using experimental data.

In general, however, the scale dependences of $\Psi^{(\delta\chi)}$
and $S_{\phi\chi}^{(\delta\phi)}$ are different.  In the slow-roll
inflation scenario, $\Psi^{(\delta\chi)}$ is generated from the
fluctuation of the inflaton amplitude and is given by \cite{PRD28-629}
\begin{eqnarray}
\Psi^{(\delta\chi)}_{\rm RD2} = \frac{4}{9}
\left[ \frac{H_{\rm inf}}{2\pi}\frac{3H_{\rm inf}^2}{V_{\rm inf}'} 
\right]_{k=aH_{\rm inf}},
\label{Psi(dchi)_RD2}
\end{eqnarray}
where $V_{\rm inf}'\equiv\partial V_{\rm inf}/\partial\chi$ with
$V_{\rm inf}$ being the inflaton potential.  On the contrary,
$S_{\phi\chi}^{(\delta\phi)}$ is related to the fluctuation of the
amplitude of $\phi$, as seen in Eq.\ (\ref{s_pc}), and hence
\begin{eqnarray}
S_{\phi\chi}^{(\delta\phi)} = 
\frac{2}{\phi_{\rm init}} 
\left[ \frac{H_{\rm inf}}{2\pi} \right]_{k=aH_{\rm inf}}.
\label{S(dphi)}
\end{eqnarray}
In many models of slow-roll inflation, the expansion rate $H_{\rm
inf}$ is almost constant during the inflation.  On the contrary, the
slope of the inflation potential $V_{\rm inf}'$ may significantly
vary.  As a result, $S_{\phi\chi}^{(\delta\phi)}$ becomes (almost)
scale independent while $\Psi^{(\delta\chi)}$ may have sizable scale
dependence.  Since the currently measured CMB power spectrum suggests
(almost) scale invariant primordial density perturbation in the
conventional scenario, inflation models are excluded if
$\Psi^{(\delta\chi)}$ has too strong scale dependence
\cite{wang_et_al_2001}.

If the $\phi$ field exists, however, the situation may change.  Since
the variable $S_{\phi\chi}^{(\delta\phi)}$ is expected to be (almost)
scale invariant, we can relax the constraint on the inflation models
if $C_l^{(\delta\phi)}$ becomes significantly large, which happens
when $S_{\phi\chi}^{(\delta\phi)}\gtrsim\Psi^{(\delta\chi)}$.  As
shown in Eq.\ (\ref{s_pc}), $S_{\phi\chi}^{(\delta\phi)}$ is inversely
proportional to $\phi_{\rm init}$.  Thus, if the initial amplitude of
$\phi$ is suppressed, this may happen. In particular, if
$C_l^{(\delta\phi)}\gg C_l^{(\delta\chi)}$, the present cosmic density
fluctuations are totally from the primordial fluctuation of $\phi$
\cite{NPB626-395,PLB524-5,PLB522-215}.  (Such a scenario is sometimes
called ``curvaton scenario.''\footnote
{For recent discussions on the curvaton scenario, see
\cite{recent-curvaton}.})

In addition, this mechanism of converting the isocurvature density
perturbation in the scalar field to the adiabatic density perturbation
may have some significance in the pre-big-bang \cite{PreBigBang} and
ekpyrotic \cite{ekpyrotic} scenario.  In Refs.\ 
\cite{NPB626-395,PLB524-5}, it was discussed that the present
cosmological density perturbations may be generated from the
primordial fluctuation of some axion-like fields in the pre-big-bang
and ekpyrotic scenarios.

\section{Cases with Entropy Perturbations}
\label{sec:entropy}
\setcounter{equation}{0}

In this section, we discuss the CMB angular power spectrum in cases
with entropy perturbations.  In particular, we study the effects of
the correlated entropy perturbation generated in the cases where the
baryonic or the CDM component is generated from $\psi$, not from
$\phi$.  As we mentioned, if the $\psi$ field starts to oscillate
before the $\phi$D epoch starts, the relation (\ref{Sb}), (\ref{Sc})
or (\ref{Sbc}) holds.  On the contrary, if $\psi$ is slow-rolling at
the time of the $\phi$-domination, density perturbations become
adiabatic.  Thus, in this section, we consider the most typical
relations (\ref{Sb}) $-$ (\ref{Sbc}) when the adiabatic and
isocurvature perturbations exist with cross-correlation.  In the
general case, as we mentioned, the relation between the adiabatic and
isocurvature perturbations may deviate from Eqs.\ (\ref{Sb}) $-$
(\ref{Sbc}).  This may happen, however, only if the expansion rate at
the beginning of the $\phi$D epoch becomes accidentally close to
$m_\psi$, and we will not consider such a case in this paper.  (See
\cite{PRD62-043504} for such a case.)

As discussed in the previous section, the CMB power spectrum is given
in the form
\begin{eqnarray}
C_l 
= C_l^{(\delta\chi)} + C_l^{(\delta\phi)} + C_l^{(\delta\psi)}.
\end{eqnarray}
Here, $C_l^{(\delta\chi)}$ is from the inflaton fluctuation which is
usually parameterized by the metric perturbation generated by the
inflaton fluctuation (in the RD2 epoch), $\Psi_{\rm
RD2}^{(\delta\chi)}$.  In addition, $C_l^{(\delta\phi)}$ and
$C_l^{(\delta\psi)}$ are parameterized by the entropy perturbations in
the baryon or the CDM.  To parameterize these contributions, we define
the ratios
\begin{eqnarray}
R_b \equiv 
S_{b\gamma}^{(\delta\phi)} / \Psi_{\rm RD2}^{(\delta\chi)},~~~
R_c \equiv 
S_{c\gamma}^{(\delta\phi)} / \Psi_{\rm RD2}^{(\delta\chi)},
\end{eqnarray}
and 
\begin{eqnarray}
\alpha_b \equiv 
S_{b\gamma}^{(\delta\psi)} / \Psi_{\rm RD2}^{(\delta\chi)},~~~
\alpha_c \equiv 
S_{c\gamma}^{(\delta\psi)} / \Psi_{\rm RD2}^{(\delta\chi)},
\end{eqnarray}
where $S_{b\gamma}^{(\delta \phi)}$ ($S_{c\gamma}^{(\delta \phi)}$) is
the entropy between the baryon and the photon (between the CDM and the
photon) generated from the primordial fluctuation of $\phi$, and
$S_{b\gamma}^{(\delta \psi)}$ and $S_{c\gamma}^{(\delta \psi)}$ are
those from $\delta\psi$.  (We adopt Eqs.\ (\ref{Sb}) and (\ref{Sc}),
and hence $S_{b\gamma}^{(\delta\psi)}$ and
$S_{c\gamma}^{(\delta\psi)}$ are equal to $\frac{9}{2}\Psi_{\rm
RD2}^{(\delta\phi)}$ if they are non-vanishing.)  The shape of the
CMB angular power spectrum depends on the values of these parameters.

The expressions for $\Psi_{\rm RD2}^{(\delta\chi)}$ and
$S_{\phi\chi}^{(\delta\phi)}$ are given in Eqs.\ (\ref{Psi(dchi)_RD2})
and (\ref{S(dphi)}), respectively, and hence, if $\psi$ starts to
oscillate before the $\phi$D epoch starts, the $R$-parameters are
given as
\begin{eqnarray}
    R_{b,c} = \frac{3}{2} 
    \left[ \frac{V'_{\rm inf}}{\phi_{\rm init}H_{\rm inf}^2} 
    \right]_{k=aH_{\rm inf}}.
\end{eqnarray}
Hence $R_{b,c}$ is model- and scenario-dependent; it depends on the
scale of inflation, shape of the inflaton potential, and initial
amplitude of $\phi$.  For example in the chaotic inflation model with
the parabolic potential $V_{\rm inf}=\frac{1}{2}m_\chi^2\chi^2$, the
above expression becomes
\begin{eqnarray}
\left[ R_{b,c} \right]_{\rm chaotic} = 
\left[ \frac{9M_*^2}{\phi_{\rm init}\chi} \right]_{k=aH_{\rm inf}}.
\end{eqnarray}
Using the fact that the inflaton amplitude at the time of the horizon
crossing of the COBE scale is $\chi\simeq 15M_*$ in the chaotic
inflation model, $\left[R_{b,c}\right]_{\rm chaotic}\simeq
0.6M_*/\phi_{\rm init}$.  Of course, the values of $R_b$ and $R_c$
depend on the model of inflation, and they vary if we consider
different class of inflation models.

The $\alpha$-parameters are also estimated in the similar way, and the
result is given from the expression for the $R$-parameters by
replacing $\phi\rightarrow\psi$.

\subsection{Uncorrelated Entropy Perturbation}

Although our main subject is to investigate effects of the fluctuation
of the $\phi$ field, it is instructive to study effects of the
uncorrelated isocurvature perturbations.  Thus, let us first consider
the case with uncorrelated isocurvature perturbations (i.e., case with
$R_b=R_c=0$).

In this case, the resultant CMB power spectrum is given in the form
\begin{eqnarray}
C_l 
= C_l^{(\delta\chi)} 
+ \alpha^2 \left[ C_l^{(\delta\psi)} \right]_{\alpha=1},
\end{eqnarray}
where $\alpha$ here is $\alpha_b$ or $\alpha_c$ depending on the
scenario.  

\begin{figure}[t]
    \begin{center}
        \scalebox{0.75}{\includegraphics{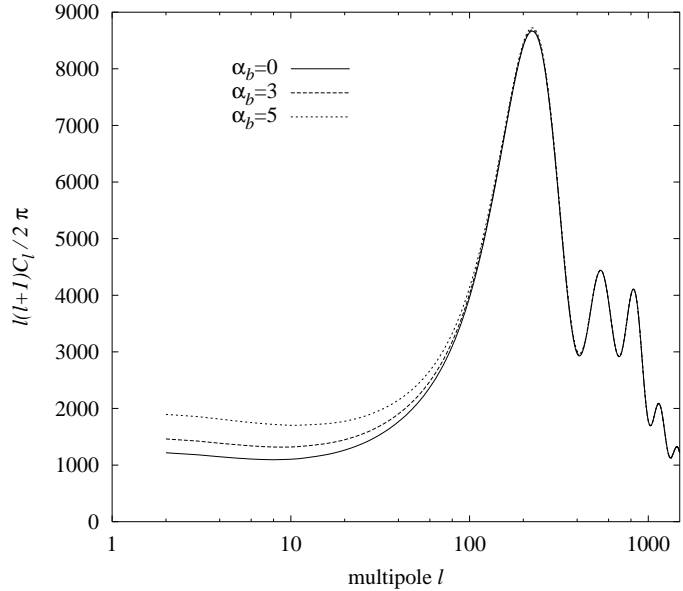}}
        \caption{The CMB angular power spectrum with the mixed 
          baryonic isocurvature and adiabatic density perturbations
          with $\alpha_b=0$ (solid line), 3 (long-dashed line) and 5
          (short-dashed line).  The cosmological parameters are the
          same as those in Fig.\ \ref{fig:Cl's}.  The overall
          normalizations are arbitrary.}
        \label{fig:Cl(iso)}
    \end{center}
\end{figure} 

In Fig.\ \ref{fig:Cl(iso)}, we plot the total angular power spectrum
for several values of $\alpha_b$ (with $\alpha_c=0$).  As can be seen,
as the effects of the isocurvature mode become more significant, the
power spectrum at the high multipole is more suppressed relative to
$C_l$ at small $l$.  In addition, since the effects of
$C_l^{(\delta\psi)}$ at high multipole are very small, the structure
of the acoustic peaks are almost unchanged from the adiabatic result
unless the $\alpha_b$-parameter becomes extremely large.

Since the isocurvature mode changes the shape of the CMB power
spectrum, upper bounds on the $\alpha$-parameters can be obtained; if
the $\alpha$-parameters are too large, the acoustic peaks are so
suppressed that the CMB power spectrum becomes inconsistent with the
observations.\footnote
{For other discussions on the constraints on the uncorrelated
isocurvature perturbations, see \cite{uncorr-iso}.}

\begin{figure}[t]
    \begin{center}
        \scalebox{0.75}{\includegraphics{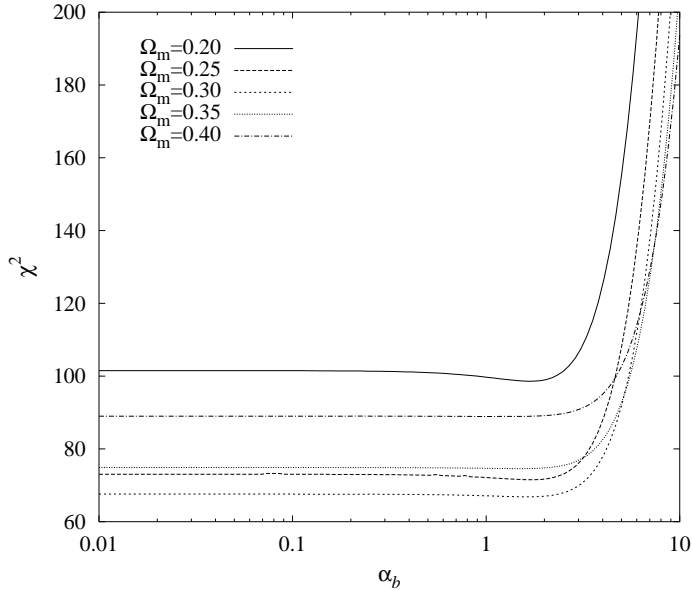}}
        \caption{$\chi^2$ as a function of $\alpha_b$.  Here we 
        take $\Omega_bh^2 = 0.019$, $h=0.65$, and the values of
        $\Omega_m$ are shown in the figure.  The flat universe is
        assumed.}
        \label{fig:chi2(iso)}
    \end{center}
\end{figure} 

As in the previous case, we calculate the $\chi^2$ variable as a
function of $\alpha_b$, and the results are shown in Fig.\ 
\ref{fig:chi2(iso)}.  From the figure, we see that too large
$\alpha_b$ is disfavored by the current observations; taking
$\Omega_m=0.3$, which gives the most conservative upper bound on
$\alpha_b$, the constraint $\alpha_b\leq 4.5$ at 95 \% C.L. is
obtained.

The CMB angular power spectrum with $\alpha_c\neq 0$ can be also
studied.  In fact, the above results with $\alpha_b\neq 0$ can be
translated to the CDM isocurvature cases; the CMB angular power
spectrum for the baryonic and the CDM isocurvature cases are the same
if $\Omega_b\alpha_b=\Omega_c\alpha_c$; using this relation, we can
also derive the upper bound on $\alpha_c$.  Thus, in any case, the
heights of the acoustic peaks decrease relative to the SW tail as the
(uncorrelated) isocurvature fluctuation contaminates.

\subsection{Correlated Entropy Perturbation in the Baryonic Sector}

Now, we study effects of the correlated entropy fluctuation in the
baryonic sector (i.e., case with $R_b\neq 0$).  Here, we assume other
entropy perturbations to vanish (i.e., $R_c=\alpha_b=\alpha_c=0$).
The general case will be discussed later.\footnote
{Notice that the $\alpha$-parameters vanish, for example, for the case
where the effective mass of the $\psi$ field becomes comparable to (or
larger than) the expansion rate during the inflation.}
In this case, the
total angular power spectrum is given in the form
\begin{eqnarray}
C_l 
= C_l^{(\delta\chi)} 
+ R_b^2 \left[ C_l^{(\delta\phi)} \right]_{R_b=1}.
\end{eqnarray}
Thus, we expect that, as the $R_b$-parameter increases, the angular
power spectrum at the high multipole is more enhanced relative to
$C_l$ at low multipole.

\begin{figure}[t]
    \begin{center}
        \scalebox{0.75}{\includegraphics{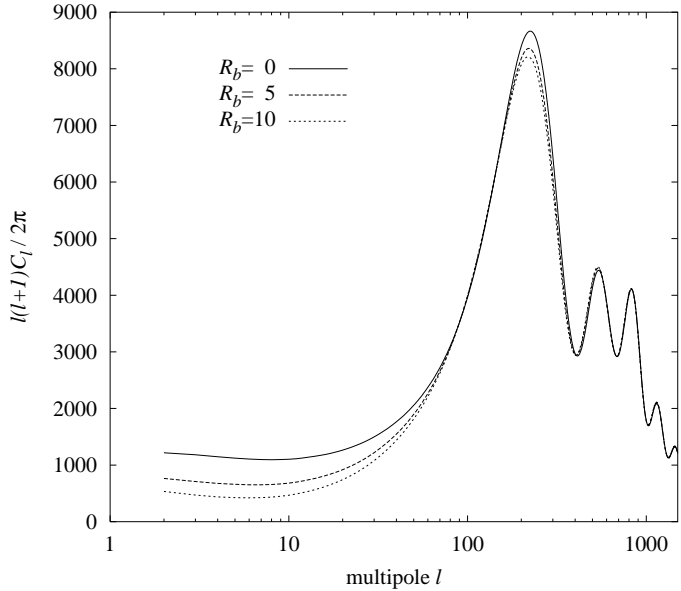}}
        \caption{The CMB angular power spectrum for the case 
        of $\alpha_b=0$.  Here we take $R_b=0$ (solid line), 5
        (long-dashed line) and 10 (short-dashed line).  The
        cosmological parameters are the same as those in Fig.\ 
        \ref{fig:Cl's}.  The overall normalizations are arbitrary.}
        \label{fig:Cl(phi)}
    \end{center}
\end{figure} 

In Fig.\ \ref{fig:Cl(phi)}, we plot the resultant angular power
spectrum with several values of $R_b$.  As expected, $C_l$ at the high
multipole is more enhanced relative to that at low ones as the
$R_b$-parameter increases, and hence it can be a signal of the
late-decaying scalar condensation.

\begin{figure}[t]
    \begin{center}
        \scalebox{0.75}{\includegraphics{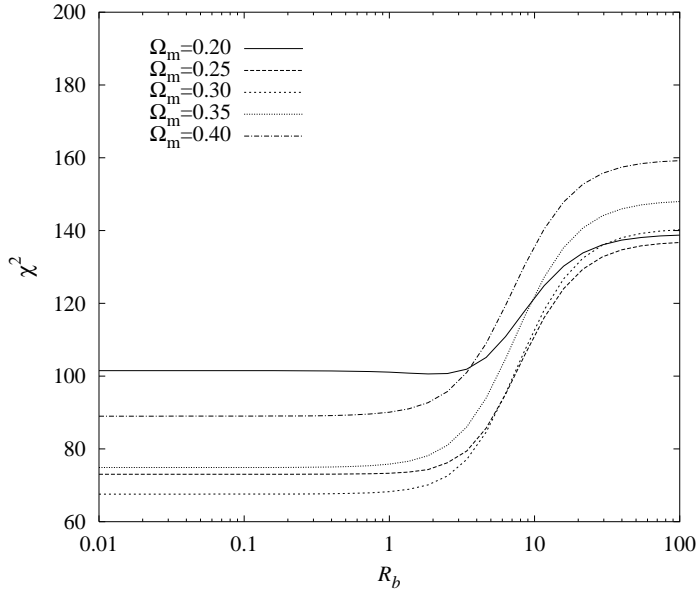}}
        \caption{$\chi^2$ as a function of $R_b$.  We take 
        $\Omega_bh^2=0.019$ and $h=0.65$. The values of $\Omega_m$ are
        shown in the figures.  The flat universe is assumed.}
        \label{fig:chi2(Rb)}
    \end{center}
\end{figure} 

In fact, if the $R_b$-parameter is too large, the angular power
spectrum at high multipole is too enhanced, which becomes
inconsistent with the currently available data.  For the quantitative
discussion, we calculate the goodness-of-fit parameter as the previous
case as a function of $R_b$.  The results are shown in Fig.\ 
\ref{fig:chi2(Rb)} for several sets of the cosmological parameters.
The figure shows that too large $R_b$ becomes inconsistent with the
observations {\sl if the total angular power spectrum is given by the
  sum of $C_l^{(\delta\chi)}$ and $C_l^{(\delta\phi)}$}.  Numerically,
even adopting the most conservative constraint, $R_b\geq 4.5$ is
excluded at 95 \% C.L.

Before closing this subsection, we would like to comment on the case
where the cosmic density perturbations are totally from the primordial
fluctuation of $\phi$, i.e., the case of the curvaton scenario.  In
this class of scenario, if the curvaton field $\phi$ does not generate
the baryon (or the CDM), the adiabatic and isocurvature perturbations
may be generated with cross-correlation with the relation given in
Eq.\ (\ref{Sb}) (or Eq.\ (\ref{Sc})).\footnote
{In more complicated cases, this may not be the case.  If some
fraction of the baryon is from the decay product of $\phi$ while the
rest is from $\psi$, for example, the relation given in Eq.\ 
(\ref{Sb}) does not hold.  In this paper, however, we do not consider
such a complicated scenario.}
The curvaton scenario requires $C_l^{(\delta\phi)}\gg
C_l^{(\delta\chi)}$ and hence $R_b$ (or $R_c$) becomes much larger
than 1, which is already excluded by the current observations.
($R_c\gg 1$ is also excluded as will be shown in the next subsection.)
One caveat to this argument is that, if the $\psi$ field is so light
that it starts to oscillate after the curvaton field dominates the
universe, $R_b$ (or $R_c$) vanishes and such a scenario is consistent
with the current data.

\subsection{Correlated Entropy Perturbation in the CDM Sector}

\begin{figure}[t]
    \begin{center}
        \scalebox{0.75}{\includegraphics{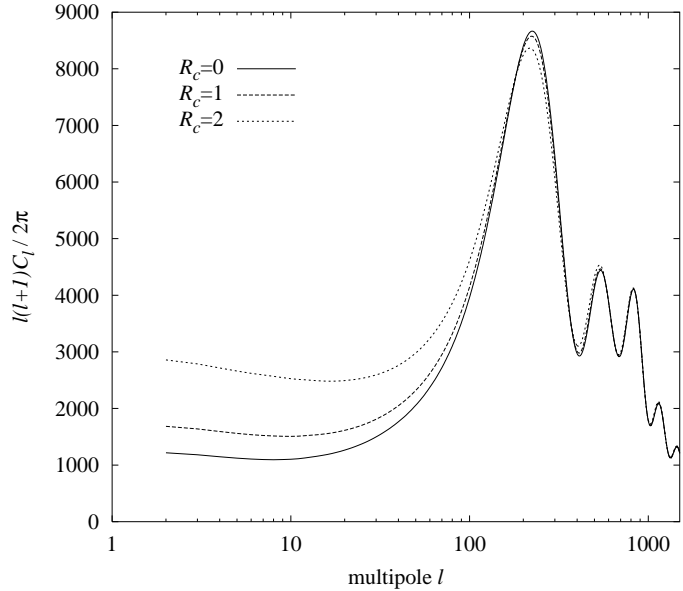}}
        \caption{The CMB angular power spectrum for the case 
        with correlated entropy perturbation in the CDM sector with
        $R_c=0$ (solid line), 1 (long-dashed line), and 2
        (short-dashed line).  The cosmological parameters are the same
        as Fig.\ \ref{fig:Cl's}.  The overall normalization are
        arbitrary.}
        \label{fig:cl(Rc)}
    \end{center}
\end{figure} 

In this subsection, we consider the case with the correlated entropy
perturbation in the CDM sector.  Effects of the correlated entropy
perturbation in the CDM sector is different from those in the baryonic
sector.  Since the density parameter for the CDM $\Omega_c$ is
significantly larger than $\Omega_b$, entropy between the
non-relativistic matter and the radiation becomes larger compared to
the baryonic case.  Then, the effects of the correlation between the
adiabatic and isocurvature perturbations become less significant while
the effect of the isocurvature perturbation itself becomes more
important.  As a result, as discussed in the previous section, if
there exists the correlated entropy perturbation in the CDM sector
with the relation (\ref{Sc}), the acoustic peaks become lower relative
to the SW tail.  In Fig.\ \ref{fig:cl(Rc)}, we show the resultant CMB
angular power spectrum for several values of $R_c$ (with
$R_b=\alpha_b=\alpha_c=0$).  As $R_c$ increases, the acoustic peaks
are more suppressed relative to the $C_l$ at lower multipole.

\begin{figure}[t]
    \begin{center}
        \scalebox{0.75}{\includegraphics{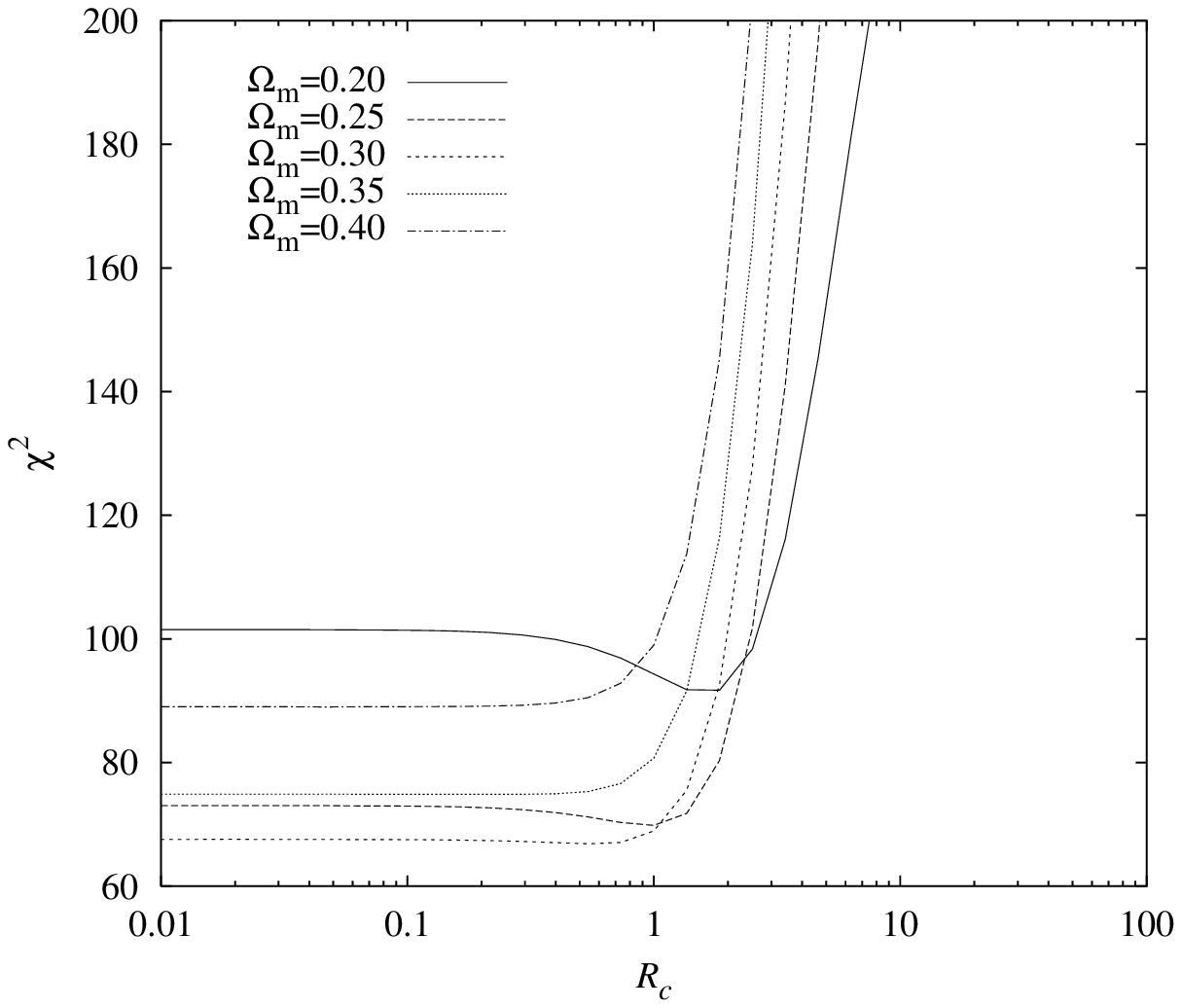}}
        \caption{$\chi^2$ as a function of $R_c$.  We take 
        $\Omega_bh^2=0.019$, $h=0.65$, and the values of $\Omega_m$
        are shown in the figure.  The flat universe is assumed.}
        \label{fig:chi2(Rc)}
    \end{center}
\end{figure} 

As in the previous cases, if the effect of the isocurvature
perturbation becomes too effective, the resultant power spectrum
becomes inconsistent with the observations.  Thus, using the current
experimental data, we can put an upper bound on the $R_c$-parameter.
In Fig.\ \ref{fig:chi2(Rc)}, we plot the goodness-of-fit parameter
$\chi^2$ as a function of $R_c$ for several sets of the cosmological
parameters.  As one can see, even adopting the most conservative case,
$R_c$ larger than 2.0 is excluded at 95 \% C.L.

\subsection{General Case}

Now, we are ready to study the most general case, the case with $R\neq
0$ and $\alpha\neq 0$.  In this case, the total angular power spectrum
is written in the form
\begin{eqnarray}
    C_l = 
    C_l^{(\delta \chi)} 
    +      R^2 \left[C_l^{(\delta\phi)} \right]_{R=1}
    + \alpha^2 \left[C_l^{(\delta\chi)} \right]_{\alpha=1}.
\end{eqnarray}
Effects of the non-vanishing $R$- and $\alpha$-parameters can be read
off from Fig.\ \ref{fig:Cl's}; normalizing the SW tail, non-vanishing
$R_b$ increases the heights of the acoustic peaks while non-vanishing
$R_c$, $\alpha_b$, and $\alpha_c$ suppress the acoustic peaks.

\begin{figure}
    \begin{center}
        \scalebox{1.1}{\includegraphics{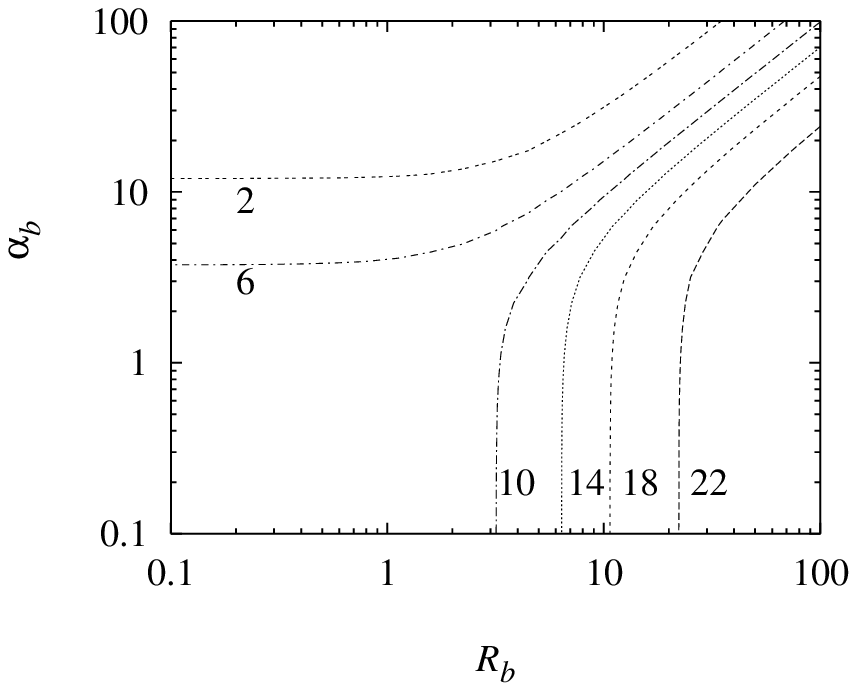}}
        \caption{Contours of constant $C_{\rm 1st}/C_{10}$ on the 
        $R_b$ vs.\ $\alpha_b$ plane.  (The ratios are shown in the
        figure.)  Here, $R_c=\alpha_c=0$ and we take the cosmological
        parameters same as Fig.\ \ref{fig:Cl's}.}
        \label{fig:c1st}
    \end{center}
%
    \begin{center}
        \scalebox{1.1}{\includegraphics{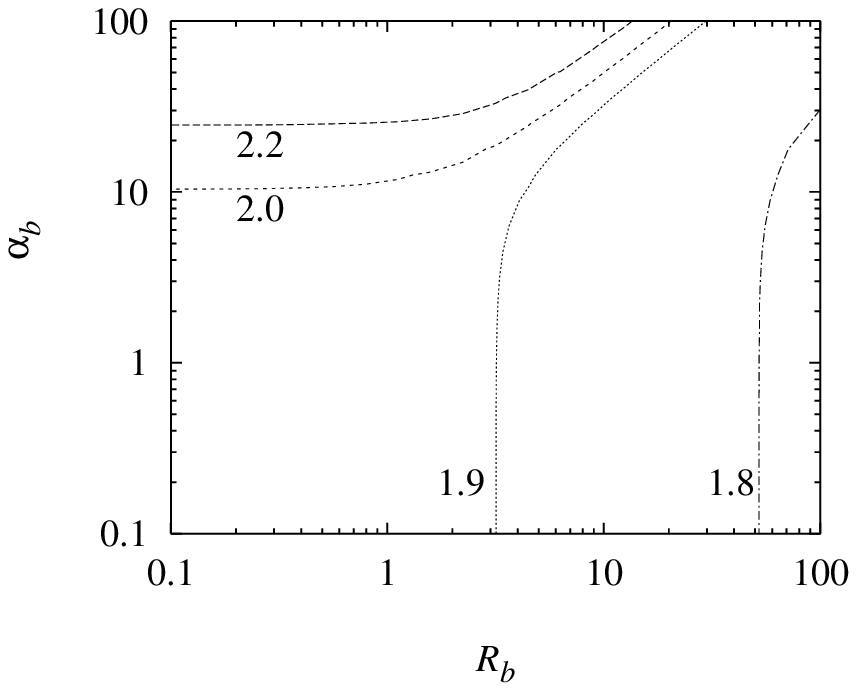}}
        \caption{Same as Fig.\ \ref{fig:c1st}, except that the
        contours are those of constant $C_{\rm 1st}/C_{\rm 2nd}$.}
        \label{fig:c2nd}
    \end{center}
\end{figure} 

First, let us consider the baryonic isocurvature perturbations.  In
Figs.\ \ref{fig:c1st} and \ref{fig:c2nd}, we show the contours of the
constant $C_{\rm 1st}/C_{10}$ and $C_{\rm 1st}/C_{\rm 2nd}$ on the
$R_b$ vs.\ $\alpha_b$ plane, where $C_{\rm 1st}$ and $C_{\rm 2nd}$ are
the heights of the first and second acoustic peaks, respectively.  As
one can see, larger value of $R_b$ ($\alpha_b$) corresponds to larger
(smaller) ratio of $C_{\rm 1st}/C_{10}$.  Importantly, the correlated
and uncorrelated entropy perturbations affect the heights of the
acoustic peaks in the opposite way.  Thus, even with extremely large
values of the $R_b$- and $\alpha_b$-parameters, the ratio $C_{\rm
1st}/C_{10}$ may take a similar value as that from the adiabatic case
if $R_b$ and $\alpha_b$ are properly related.  The ratio $C_{\rm
1st}/C_{\rm 2nd}$ is less sensitive to the parameters $R_b$ and
$\alpha_b$, and for larger (smaller) value of $R_b$ ($\alpha_b$), the
ratio $C_{\rm 1st}/C_{\rm 2nd}$ is suppressed.  We also calculated the
ratio $C_{\rm 3rd}/C_{\rm 2nd}$ (with $C_{\rm 3rd}$ being the height
of the third acoustic peak) and found that this ratio is almost
independent of $R_b$ and $\alpha_b$.  This is because the structure of
the angular power spectrum at high multipole is dominantly determined
by the adiabatic part of the perturnations.  We would like to
emphasize that the effects of the primordial fluctuations of the
$\phi$ and $\psi$ field are quite different; $C_l^{(\delta\phi)}$ is
enhanced at high multipole while $C_l^{(\delta\psi)}$ becomes larger
at low one.  Thus, if the parameters $R_b$ and $\alpha_b$ obey a
specific relation, the adiabatic result may be mimiced even when the
effects of the entropy perturbation are quite large.

\begin{figure}
    \begin{center}
        \scalebox{0.5}{
        \includegraphics{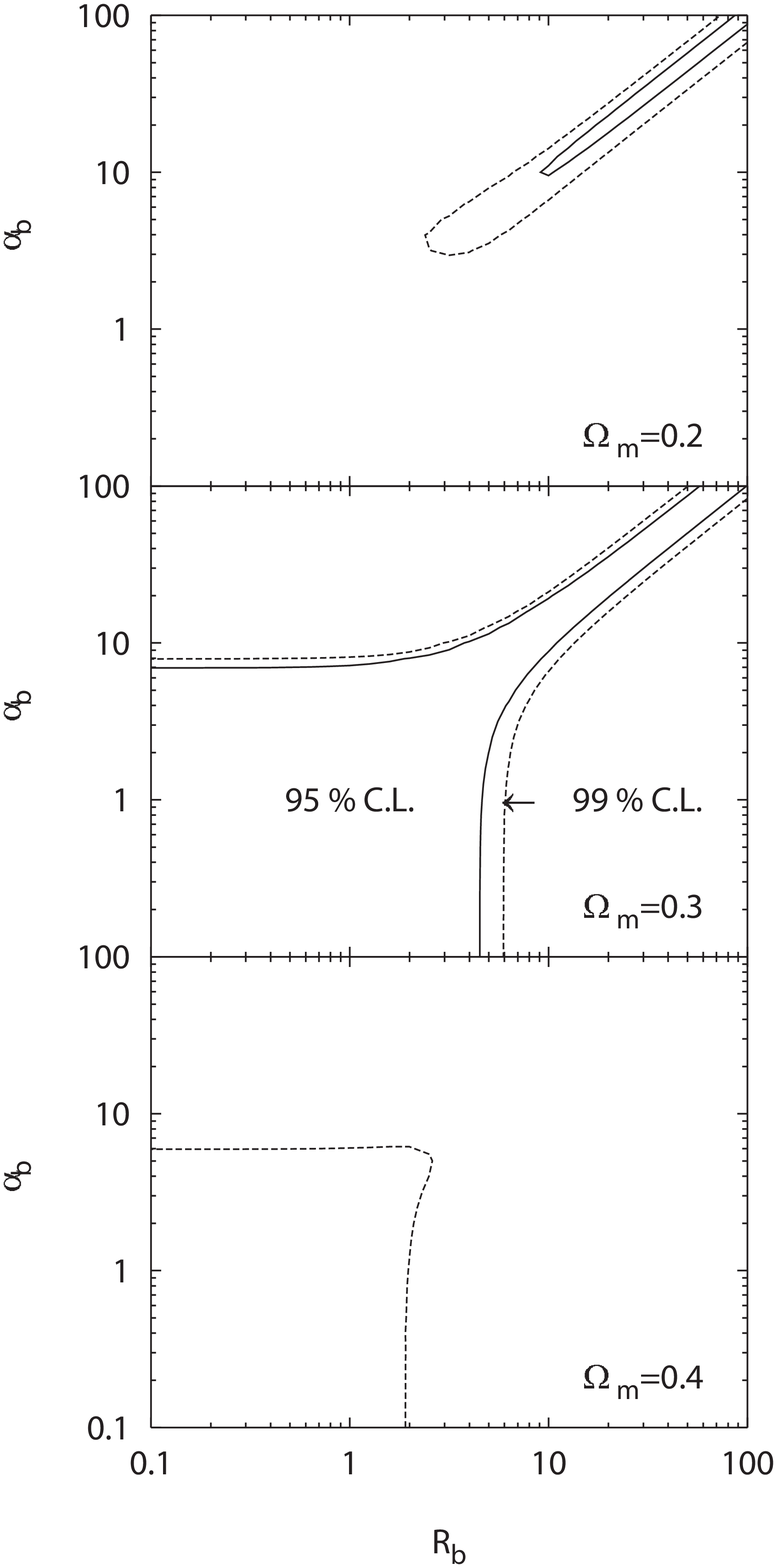}
        }
        \caption{Contours of constant $\chi^2$ on the $R_b$ vs.\ 
        $\alpha_b$ plane.  We show contours of $\chi^2=84$ (solid
        line) and 93 (dashed line), which correspond to 95 \% and 99
        \% C.L. allowed region with the $\chi^2$ statistics with 64
        degrees of freedom, respectively.  Here we take
        $\Omega_bh^2=0.019$, $h=0.65$, and the values of $\Omega_m$
        are shown in the figures.}
        \label{fig:chi2(Ra)}
    \end{center}
\end{figure}  

To study this issue, we calculate the goodness-of-fit parameter
$\chi^2$ as a function of $R_b$ and $\alpha_b$ (with
$R_c=\alpha_c=0$).  Here, the overall normalization of $C_l$ is chosen
such that the $\chi^2$ variable is minimized.  The results are shown
in Fig.\ \ref{fig:chi2(Ra)} for several sets of the cosmological
parameters.  Here, all the primordial density perturbations are
assumed to be scale-invariant.  For the cases with $\Omega_m=0.2$ and
0.3, it is interesting that large values of $R_b$ and $\alpha_b$ are
allowed.  In particular, for the case with $\Omega_m=0.2$, only the
case with non-vanishing $R_b$ and $\alpha_b$ is allowed. Importantly,
in this parameter region, the CMB anisotropy originates to
non-adiabatic density perturbations while the contribution from the
inflaton fluctuation $C_l^{(\delta\chi)}$ becomes negligibly small.
In the following, we call this case as ``modified-curvaton'' case.  It
should be noticed that, in the modified-curvaton case, the parameters
$R_b$ and $\alpha_b$ satisfy the relation $\alpha_b\simeq
(1.0-1.7)R_b$ for $\Omega_m=0.3$.  For $\Omega_m=0.4$, large values of
$R_b$ and $\alpha_b$ are not allowed in contrast to other cases.

\begin{figure}[t]
    \begin{center}
        \scalebox{0.75}{\includegraphics{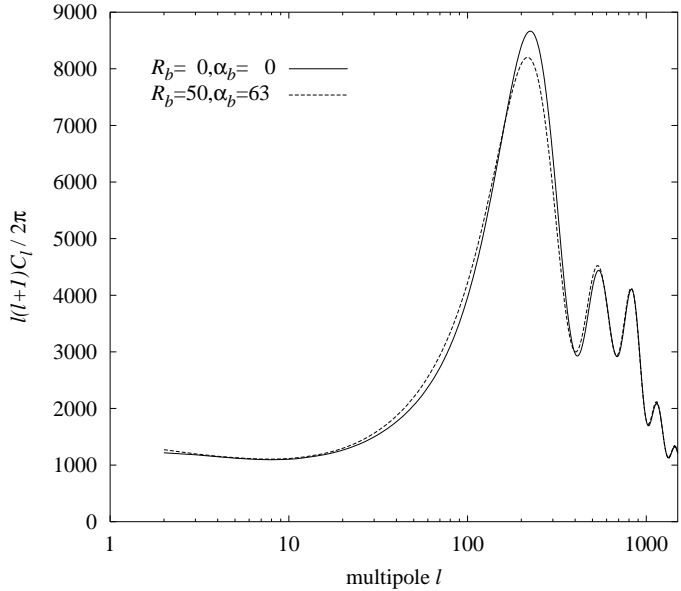}}
        \caption{The angular power spectrum with $R_b=\alpha_b=0$ 
        (solid line), and $R_b=50$ and $\alpha_b=63$ (dashed line).
        We take $R_c=\alpha_c=0$.  The cosmological parameters are the
        same as those in Fig.\ \ref{fig:Cl's}.  The overall
        normalizations are arbitrary.}
        \label{fig:cl(adiiso)}
    \end{center}
\end{figure} 

It is interesting to study whether the angular power spectrum in the
modified-curvaton case can be distinguished from the adiabatic one by
the on-going and future experiments.  In Fig.\ \ref{fig:cl(adiiso)},
we plot the CMB angular power spectrum in the adiabatic and
modified-curvaton cases, $R_b=\alpha_b=0$, and $R_b=50$ and
$\alpha_b=63$ (which minimizes the $\chi^2$ variable for $R_b=50$),
respectively.  In the modified-curvaton case, shape of the power
spectrum at high multipole is determined by $C_l^{(\delta\phi)}$ which
is similar to that of the adiabatic case.  Thus, it is difficult to
distinguish two cases just by looking at the power spectrum at high
multipole.  However, as one can see, the structure of the angular
power spectrum at relatively low multipole is significantly
affected. In particular, the ratio $C_{\rm 1st}/C_{\rm 2nd}$ will
provide us an interesting probe for studying the modified-curvaton
scenario.  Since the MAP is expected to measure the angular power
spectrum at $O(1\ \%)$ level up to the multipole $l\sim 1000 - 2000$,
$C_l$ at the high multipole may be used to calibrate the power
spectrum to see the distortion of the CMB angular power spectrum from
the adiabatic result.

We can also consider the case where the CDM sector has correlated and
uncorrelated entropy perturbations (i.e., the case with $R_c\neq 0$
and $\alpha_c\neq 0$).  Importantly, correlated and uncorrelated
entropy perturbations both suppress the angular power spectrum at high
multipole relative to the low one.  Thus, in this case, too large
$R_c$ is always excluded irrespective of the value of $\alpha_c$ if
the baryon is from the decay product of $\phi$.

\section{Conclusion}
\label{sec:conclusion}
\setcounter{equation}{0}

In this paper, we discussed the effects of the late-time entropy
production due to the decay of the scalar-field condensations on the
cosmic density perturbations.  If the universe is reheated by the
decay of the scalar field $\phi$, many of the components in the
present universe are generated from the decay products of the $\phi$
field.  In such a case, cosmic density perturbations are affected by
the fluctuation of the amplitude of $\phi$ which may be generated
during the inflation.

If all the components in the universe originate to the decay product
of $\phi$, density perturbations generated from the primordial
fluctuation of $\phi$ becomes adiabatic.  In this case, the CMB
angular power spectrum from the fluctuation of $\phi$ becomes the
usual adiabatic ones with (almost) scale-invariant spectrum.  If this
becomes the dominant part of the cosmic density perturbations, then we
have seen that the constraints on inflation models from 
observations of the CMB angular power spectrum are drastically relaxed.

If the baryon or the CDM is not generated from $\phi$ but from a new
scalar field $\psi$, on the contrary, correlated mixture of the
adiabatic and isocurvature perturbations may arise.  In particular, if
the $\psi$ field starts to oscillate much before the $\phi$ field
dominates the universe, the metric and entropy perturbations obey the
model-independent relation.  In this case, the CMB angular power
spectrum may be significantly affected and the shape of the resultant
power spectrum depends on which component has the correlated
isocurvature perturbation.  If the baryonic component has the
correlated isocurvature perturbation, the density perturbations after
the decay of $\phi$ become those given in Eq.\ (\ref{Sb}).  In this
case, $C_l$ at high multipole is more enhanced relative to that at low
multipole.  On the contrary, if the CDM is not generated from $\phi$,
correlated isocurvature perturbation can be induced in the CDM
component as given in Eq.\ (\ref{Sc}).  In this case, heights of the
acoustic peaks become lower relative to the SW tail.  If there exists
an contamination of these component into the total angular power
spectrum, $C_l$ differs from the adiabatic one.  The important point
is that, in both cases, too much correlated isocurvature perturbations
become inconsistent with the present observation of the CMB power
spectrum.

Even with the correlated isocurvature perturbation in the baryonic
sector, heights of the acoustic peaks can be reduced by introducing
uncorrelated isocurvature perturbation.  This may happen, for example,
if we take account of the primordial fluctuation of the scalar field
$\psi$.  (For example, this scalar field may be the Affleck-Dine
field.)  In particular, if the sizes of the correlated and
uncorrelated entropy perturbations are properly chosen, the resultant
CMB power spectrum becomes consistent with the present observations
without the effect of the inflaton fluctuation.  Even in this case,
the angular power spectrum is not exactly the same as the conventional
adiabatic one and the deviation from the adiabatic result may be
observed at on-going and future experiments.

In summary, if there exists correlated mixture of the adiabatic and
isocurvature perturbations, the total CMB angular power spectrum can
be distorted and it can be a signal of the late-time entropy
production due to the decay of the scalar field condensation.  The
on-going and future experiments may observe such a signal.

{\sl Acknowledgment:} We acknowledge the use of CMBFAST \cite{cmbfast}
and RADPACK \cite{radpack} packages for our numerical calculations.
This work is supported by the Grant-in-Aid for Scientific Research
from the Ministry of Education, Science, Sports, and Culture of Japan,
No.\ 12047201 and No.\ 13740138.


\begin{thebibliography}{100}

\bibitem{APJ464L1}
    C. Bennett et al.,
    Astrophys.\ J.\ Lett.\ {\bf 464} (1996) L1.

\bibitem{aph0104460}
    C.B.\ Netterfield et al., 
    astro-ph/0104460.

\bibitem{aph0104459}
    A.T.\ Lee et al., 
    astro-ph/0104459.

\bibitem{aph0104489}
    N.W.\ Halverson et al., 
    astro-ph/0104489.

\bibitem{MAP}
    MAP webpage,
    {\tt http://map.gsfc.nasa.gov}.

\bibitem{PLANCK} 
    PLANCK webpage,
    {\tt http://astro.estec.esa.nl/SA-general/Projects/Planck}.

\bibitem{PRD23-347}
  A.H.\ Guth, 
  Phys.\ Rev.\ {\bf D23} (1981) 347.

\bibitem{hep-ph_0201264}
  See, for example, R. Liddle,
  hep-ph/0201264.

\bibitem{NPB249-361}
    I. Affleck and M. Dine,
    Nucl.\ Phys.\ {\bf B249} (1985) 361.

\bibitem{Polchinski}
    See, for example, J. Polchinski,
    ``String Theory'' (1998, Cambridge University Press).

\bibitem{PRL131-59}
    G.D. Coughlan, W. Fischler, E.W. Kolb, S.Raby
    and G.G. Ross,
    Phys.\ Lett.\ {\bf B131} (1983) 59.

\bibitem{heavy-moduli}
    T. Moroi, M. Yamaguchi and T. Yanagida,
    Phys.\ Lett.\ {\bf B342} (1995) 105;
    M. Kawasaki, T. Moroi and T. Yanagida,
    Phys.\ Lett.\ {\bf B370} (1996) 52;
    T. Moroi and L. Randall,
    Nucl.\ Phys.\ {\bf B570} (2000) 455.

\bibitem{snu-leptogen}
    H. Murayama and T. Yanagida,
    Phys.\ Lett.\ {\bf B322} (1994) 349;
    K. Hamaguchi, H. Murayama and T. Yanagida,
    Phys.\ Rev.\ {\bf D65} (2002) 043512.

\bibitem{axion=seeds}
    R. Durrer, M. Gasperini, M. Sakellariadou and G. Veneziano,
    Phys.\ Rev.\ {\bf D59} (1999) 043511;
    R. Durrer, M. Gasperini, M. Sakellariadou and G. Veneziano,
    Phys.\ Lett.\ {\bf B436} (1998) 66;
    A. Melchiorri, F. Vernizzi, R. Durrer and G. Veneziano,
    Phys.\ Rev.\ Lett.\ {\bf 83} (1999) 4464.

\bibitem{PreBigBang}
    G. Veneziano, 
    Phys.\ Lett.\ {\bf B265} (1991) 287;
    M. Gasperini and G. Veneziano, 
    Astropart.\ Phys.\ {\bf 1} (1992)  1; 
    Phys.\ Rev.\ {\bf D50} (1994) 2519.

\bibitem{ekpyrotic}
    J. Khoury, B.A. Ovrut, P.J. Steinhardt and N. Turok, 
    Phys.\ Rev.\ {\bf D64} (2001) 123522;
    J. Khoury, B.A. Ovrut, N. Seiberg, P.J. Steinhardt and N. Turok, 
    hep-th/0108187;
    J. Khoury, B.A. Ovrut, P.J. Steinhardt and N. Turok, 
    hep-th/0109050.

\bibitem{PRD42-313}
    S. Mollerach,
    Phys.\ Rev.\ {\bf D42} (1990) 313.

\bibitem{NPB626-395}
    K. Enqvist and M.S. Sloth,
    Nucl.\ Phys.\ {\bf B626} (2002) 395.

\bibitem{PLB524-5}
    D.H. Lyth and D. Wands,
    Phys.\ Lett. {\bf B524} (2002) 5.

\bibitem{PLB522-215}
    T. Moroi and T.\ Takahashi,
    Phys.\ Lett. {\bf B522} (2001) 215.
    
\bibitem{adi-iso}
    D. Langlois.
    Phys.\ Rev.\ {\bf D59} (1999) 123512;
    C. Gordon, D. Wands, B.A. Bassett and R. Maartens,
    Phys.\ Rev.\ {\bf D63} (2001) 023506;
    N. Bartolo, S. Matarrese and A. Riotto,
    Phys.\ Rev.\ {\bf D64} (2001) 083514; 
    Phys.\ Rev.\ {\bf D64} (2001) 123504;
    L. Amendola, C. Gordon, D. Wands and M. Sasaki,
    astro-ph/0107089.

\bibitem{DinRanTho}
    M. Dine, L.J. Randall and S. Thomas,
    Phys.\ Rev.\ Lett.\ {\bf 75} (1995) 398;
    Nucl.\ Phys.\ {\bf B458} (1996) 291.

\bibitem{hu_PhD_thesis}
    W. Hu,
    Ph.D thesis (astro-ph/9508126).

\bibitem{primack}
    J. Primack,
    astro-ph/0007187.

\bibitem{Bond_Jaffe_Knox_2000}
    J.R. Bond, A.H. Jaffe and L. Knox,
    Astrophys.\ J.\ {\bf 533} (2000) 19.

\bibitem{PRD28-629}
    J.M. Bardeen, P.J. Steinhardt and M.S. Turner,
    Phys.\ Rev.\ {\bf D28} (1983) 629.

\bibitem{wang_et_al_2001}
    X. Wang, M. Tegmark and M. Zaldarriaga,
    astro-ph/0105091.

\bibitem{recent-curvaton}
    N. Bartolo and A.R. Liddle,
    astro-ph/0203076; 
    R. Durrer and F. Vernizzi,
    hep-ph/0203275;
    A. Notari and A. Riotto,
    hep-th/0205019;
    D.H. Lyth,
    hep-ph/0205266.

\bibitem{PRD62-043504}
    D. Langlois and A. Riazuelo,
    Phys.\ Rev.\ {\bf D62} (2000) 043504.

\bibitem{uncorr-iso}
    R. Stompor, A.J. Banday and K.M. Gorski,
    Astrophys.\ J.\ {\bf 463}(1996) 8;
    S.D. Burns, astro-ph/9711303;
    T. Kanazawa, M. Kawasaki, N. Sugiyama and T. Yanagida,
    Prog.\ Theor.\ Phys.\ {\bf 102} (1999) 71;
    K. Enqvist, H. Kurki-Suonio and J. Valiviita,
    Phys.\ Rev.\ {\bf D62} (2000) 03003;
    Phys.\ Rev.\ {\bf D65} (2002) 043002.

\bibitem{cmbfast}
    M. Zaldarriaga and U. Seljak,
    Astrophys.\ J.\ {\bf 469} (1996) 437.

\bibitem{radpack}
    RADPACK webpage,
    {\tt http://flight.uchicago.edu/knox/radpack.html}.

\end{thebibliography}
\end{document}